\documentclass[twocolumn,english,aps,prb]{revtex4}
\usepackage{color}
\usepackage{amsmath}
\usepackage{graphicx}
\usepackage{amssymb}
\usepackage{hyperref}
\usepackage{dsfont}

\makeatletter
\@ifundefined{textcolor}{}
{%
 \definecolor{BLACK}{gray}{0}\definecolor{WHITE}{gray}{1}
 \definecolor{RED}{rgb}{1,0,0}
 \definecolor{GREEN}{rgb}{0,1,0}
 \definecolor{BLUE}{rgb}{0,0,1}
 \definecolor{CYAN}{cmyk}{1,0,0,0}
 \definecolor{MAGENTA}{cmyk}{0,1,0,0}
 \definecolor{YELLOW}{cmyk}{0,0,1,0}
 }

\@ifundefined{definecolor}
 {\usepackage{color}}{}
\@ifundefined{definecolor}{\@ifundefined{definecolor}
 {\usepackage{color}}{}
}{}

\newcommand{\ba}{\begin{eqnarray*}}
\newcommand{\ea}{\end{eqnarray*}}
\newcommand{\baa}{\begin{eqnarray}}
\newcommand{\eaa}{\end{eqnarray}}
\newcommand{\bea}{\begin{eqnarray}}
\newcommand{\eea}{\end{eqnarray}}
\newcommand{\be}{\begin{equation}}
\newcommand{\ee}{\end{equation}}

\makeatother

\usepackage{babel}

\begin{document}

\title{Finite-frequency dissipation in a driven Kondo model}

\author{Pier Paolo Baruselli}
\affiliation{International School for Advanced Studies (SISSA), Via Bonomea 265, 34136 Trieste, Italy}
\author{Erio Tosatti}
\affiliation{International School for Advanced Studies (SISSA), Via Bonomea 265, 34136 Trieste, Italy}
\affiliation{International Center for Theoretical Physics (ICTP), Strada Costiera 11, I-34014 Trieste, Italy}
\affiliation{CNR-IOM Democritos National Simulation Center, Via Bonomea 265, 34136 Trieste, Italy}


\date{\today}
\begin{abstract}
Using both a resonant level model and the time-dependent Gutzwiller approximation,  
we study the power dissipation of a localized impurity hybridized with a conduction band 
when the hybridization is periodically switched on and off.
The total dissipated energy is proportional to the Kondo temperature, with a non-trivial frequency dependence.
At low frequencies it can be well approximated by the one of a single quench,
and is obtainable analitically;
at intermediate frequencies it undergoes oscillations;
at high frequencies, after reaching its maximum, it quickly drops to zero.
This frequency-dependent energy dissipation could be relevant to systems such as irradiated quantum dots, where Kondo can be switched at very high frequencies.  
\end{abstract}

\maketitle

\section{Introduction}

In a previous paper \cite{baruselli2017} we used a simple approach to compute the mechanical energy dissipation associated with processes where one could switch on and off the Kondo effect experienced by electrons in, e.g., a tip-operated  metal-metal contacts through magnetic impurities.
That kind of mechanica manipulation can of course assumed to involve timescales much larger than the typical electronic ones, whereby we could show that an equilibrium calculation is enough to get to the Kondo dissiption result. We then used numerical renormalization group (NRG) to the problem of claculating dissipation for a single impurity Anderson model (SIAM).

Besides these tip-based setups, however, it was demonstrated that the Kondo effect in quantum dots can be selectively controlled by irradiation with an oscillating electromagnetic field generated by a laser\cite{glazman_irradiated_qd,goldstein_irradiated_qd}.
In contrast to mechanical driving, lasers can easily reach typical electronic frequencies.
Even though in such experiments no energy dissipation was measured, it should be in principle possible to do so.

In this paper we thus want to extend the treatment of the energy dissipation associated to the Kondo effect to finite frequencies,
suggesting to look at this quantity in irradiated quantum dots as a direct application of our theory;
another candidate system could be represented by cold atoms.
Once Kondo is switched on and off at very high frequency, we must explicitly treat the time dependence, and equilibrium methods such as NRG are no longer adequate.
We have chosen to study the problem using a non-interacting resonant level model (RLM), 
and,
subsequently, 
the time-tependent (TD)\cite{schiro_TDGA,lanata_TDGA,schiro2_TDGA,Fabrizio_TDGA} Gutzwiller approximation (GA)\cite {gutzwiller_1,gutzwiller_2} for a SIAM\cite{lanata_gutzwiller},
which is a mean field method.
These methods, which approximate the Kondo resonance as a Lorentzian level at the Fermi energy,
allow to study in a simple way the effect of a finite frequency on the Kondo dissipation.
To be sure, more elaborate methods would be required to study in detail the many-body physics of the time-dependent Kondo switching.
One such candidate is TD-NRG\cite{TDNRG,TDNRG2,TDNRG_costi}, which has however also been 
criticized  \cite{Rosch_TDNRG}.

Even though the energy dissipation problem is rarely considered 
-- to our knowledge only the energy transfer of an atom moving above a surface has been esplicitly addressed \cite{PhysRevB.55.2578,PhysRevB.58.2191,PhysRevB.60.5969} --
several approaches to study time-dependent phenomena in Kondo systems have been adopted in the literature.
TD-NRG itself has been employed to study the time evolution of 
an
impurity spin \cite{TDNRG2},
similarly to quantum Monte Carlo \cite{cohen_qmc} and unitary perturbation theory\cite{kehrein_unitary_pert_theory}.
Exact results have 
also
been obtained using bosonization and refermionization for the the finite frequency conductance\cite{schiller_ac},
the spin-spin correlation function\cite{lobaskin_prb,Lobaskin2006,heyl_ackondo}  
and the spin polarization in an oscillating magnetic field\cite{iwahori_pra}
at the Toulouse point.
Moreover, the dynamic charge and magnetic susceptibility have been studied using the non-crossing approximation (NCA) \cite{PhysRevB.55.2578}. In addition, 
the conductance in a QD has been addressed using perturbation theory \cite{langreth_kondo,PhysRevB.62.8154,PhysRevLett.81.4688,PhysRevLett.81.5394,PhysRevB.61.16750}, 
the equation of motion approach\cite{PhysRevLett.76.487}, 
and time-dependent NCA\cite{PhysRevLett.74.4907,PhysRevB.61.2146} (which was also used to study the thermopower  \cite{goker_thermopower_kondo}).
Finally, the Fermi-edge problem was also investigated in an oscillating electric field \cite{ng_edge}.

This paper is organized as follows.
In the first part of the paper, Sections \ref{sec2}, \ref{sec_num_res}, we will study a strictly one-body model,
i.e. a localized level hybridized with itinerant electrons, also known as resonant level model (RLM) when the on-site energy lies at the Fermi energy,
and investigate in detail the frequency dependence of the energy dissipation when switching on and off the hybridization.
In the second part we will study a SIAM using the time-dependent Gutzwiller approximation, Sections \ref{section_ga} to  \ref{section_ga_num}.
Being a mean field approach, we can relate its results to an equivalent one-body model, i.e. the RLM.
Our results show that at low frequencies both methods agree on a linear increase of the power at low frequency, and on a peak at high frequency, see Section \ref{section_disc}.
The TD-GA shows oscillations of the dissipation as a function of frequency which are likely to be spurious.
Finally, in Section \ref{sec_concl} we draw the conclusions of our work.

\section{Non-interacting Model}\label{sec2}
\begin{figure}[bt]
\includegraphics[width=0.45\textwidth]{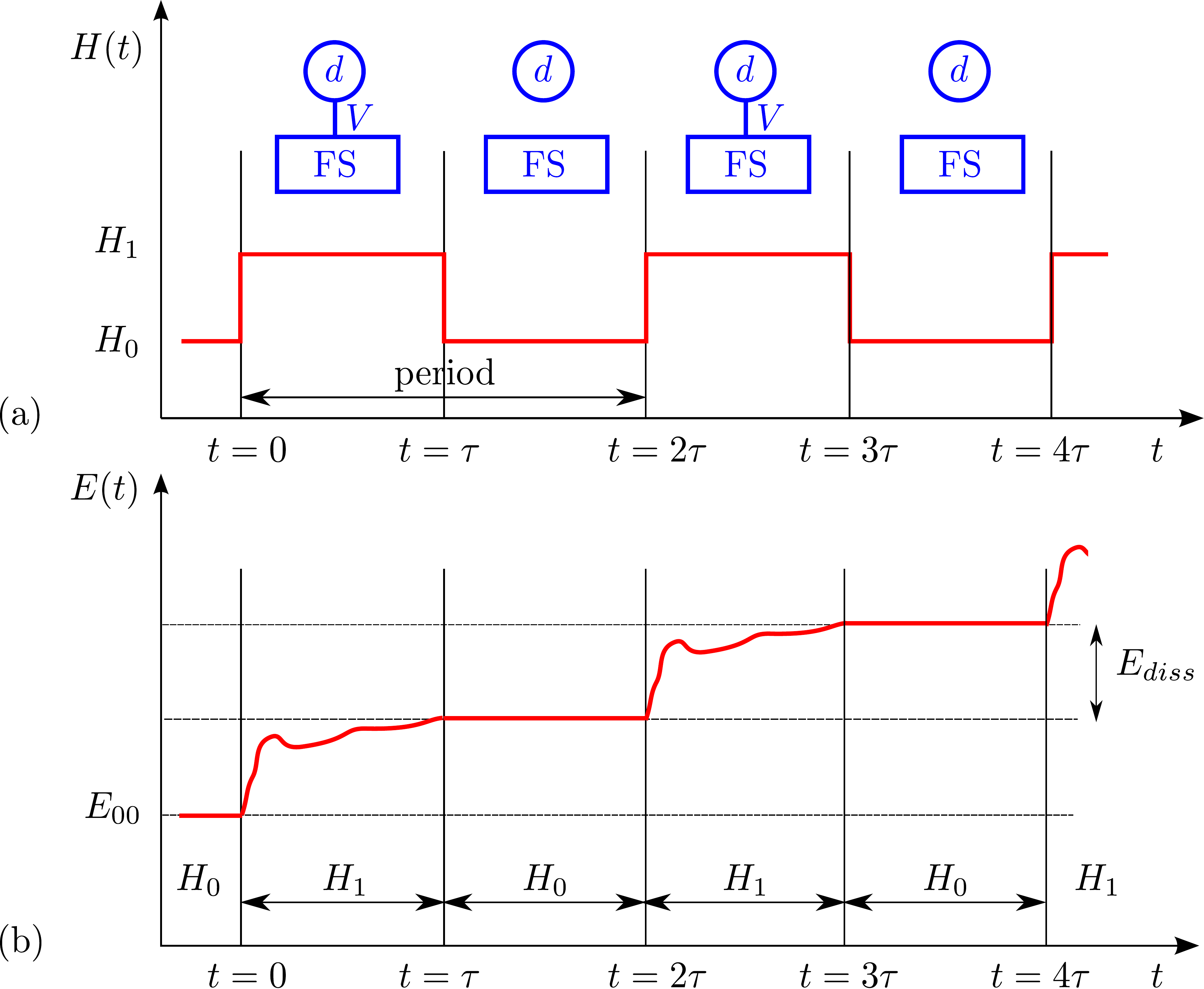}
\caption{The system  Hamiltonian is periodically modulated in a square-wave manner with period $2\tau$. In the first half of the period, the Hamiltonian is $\hat H_1$, which describes an impurity $d$ coupled to a Fermi sea (FS) by a hybridization term $V$.
In the second half of the period the hybridization is switched off, to get $\hat H_0$.
(b) Evolution of the ``internal'' energy $E_0(t)\equiv \langle \psi(t)|\hat H_0|\psi(t)\rangle$  in the non interacting model, with indication of the dissipation per cycle $E_{diss}$.
}\label{fig_H0H1}
\end{figure}

As in Ref. \onlinecite{baruselli2017}, we study a system in which the Hamiltonian can be switched from $H_0$ (no Kondo) to $H_1$ (Kondo-like) and viceversa.
We assume that the switching can be perfomed instantaneously, i.e. in a time $\tau_{switch}$ which is much smaller than all the other time scales.
We consider a periodic switching with semiperiod $\tau$ between $H_0$ and $H_1$; see Fig. \ref{fig_H0H1}.
We take as $H_0$ the Hamiltonian of a noninteracting Fermi sea (FS), plus an isolated $d$ level with energy $\epsilon_d$:
\bea
\hat H_0&=&\hat H_d+\hat T,\\
\hat H_d&=&\sum_{\sigma=\uparrow,\downarrow}\epsilon_d \hat  d_\sigma^\dagger \hat d_\sigma, \hspace{5pt} 
\hat T=\sum_{\sigma,n=-N}^{N-1} \epsilon_n \hat c_{n\sigma}^\dagger \hat c_{n\sigma} \label{eq_T} 
\eea
where we suppose that the chemical potential is set at zero energy,
with a Fermi distribution function $f(\epsilon)=(e^{\epsilon/T+1})^{-1}$ (we will work at nearly zero temperature $T=0$).
We highlight that in this resonant impurity model (unlike real Kondo) the spin index $\sigma$ does not play any important role, 
so one can work with spinless fermions, and at the end multiply all observables by a factor two.
We will consider different DOS $\rho(\epsilon)$ for the conduction states:
\be
\rho(\epsilon)=\frac{1}{2L}\sum_{n=1}^{2L} \delta(\epsilon-\epsilon_n), \hspace{5pt}\int_{-\infty}^{+\infty}\rho(\epsilon)d\epsilon=1,
\ee
namely the flat one:
\be
\rho_{flat}(\epsilon)=\frac{\theta(D-|\epsilon|)}{2D},
\ee
and the semicircular one:
\be
\rho_{sqrt}(\epsilon)=\frac{2\sqrt{D^2-\epsilon^2}\theta(D-|\epsilon|)}{\pi D}.
\ee
The final Hamiltonian $\hat H_1$ is given by $\hat H_1\equiv \hat H_0+\hat V$, where the perturbation  $\hat V \equiv \hat V_h$ is here  the hybridization between the localized level and the FS:
\bea\label{eq_v}
\hat V_h&=&\frac{V}{\sqrt{2L}}\sum_{n\sigma} (\hat c_{n\sigma}^\dagger \hat d_{\sigma}+\hat d_{\sigma}^\dagger \hat c_{n\sigma})\equiv V \hat v.   
\eea
As a consequence of  hybridization, the localized level DOS acquires a Lorentzian lineshape with broadening
$\Gamma(\epsilon_d)=\pi V^2 \rho(\epsilon_d)$;
for simplicity, in what follows we put $\Gamma\equiv\pi V^2/(2D)$.
In our numerics, we use a finite number $2L=600$ of conduction states.

\subsection{Zero frequency review}
In our previous work \cite{baruselli2017} we treated the same problem in the limit of vanishing frequencies,
which is relevant when the Kondo effect can be switched on and off mechanically,
as in an AFM experiment.
Under the hypothesis of thermalization, we found that the dissipated energy is:
\be\label{e_diss_1v1}
E_{diss}^{\tau\rightarrow\infty}=\langle \psi_0|\hat{V}|\psi_0\rangle-\langle \psi_1|\hat{V}|\psi_1\rangle,
\ee
where $|\psi_0\rangle$ is the GS of the Hamiltonian $H_0$, and $|\psi_1\rangle$ of $H_1$. 
As remarked, here $\hat V$ is the perturbation $\hat V\equiv \hat V_h$, 
so $\langle \psi_0|\hat{V}|\psi_0\rangle=0$.

For our non-interacting impurity, at $T=0$ 
the dissipation 
for $\Gamma \ll D$ is:
\bea
E_{diss}^{\tau\rightarrow\infty}&=&-\langle \psi_1|\hat{V}_h|\psi_1\rangle=\nonumber\\
&=&-\frac{2\Gamma}{\pi}\log\frac{\epsilon_d^2+\Gamma^2}{(D+|\epsilon_d|)^2+\Gamma^2},\label{e_diss_1_ex}
\eea
and acquires a $\propto \Gamma |\log T|$ behavior when $T \gg \Gamma,|\epsilon_d|$.

For a SIAM we were able to use NRG to compute this quantity, and study its temperature and magnetic field dependence.

\subsection{Finite frequencies}
In 
order to study the effect of a finite driving frequency on the dissipated energy, the TD-NRG could be in principle be implemented right away. While computationally demanding, that is also not completely trustworthy. 
Therefore, it is  wiser to start by studying a non-interacting impurity model, as was done in Ref. \onlinecite{baruselli2017} and only in the second part of the paper to address the problem by using a mean-field approach, that is, the TD-Gutzwiller approach.


For the non-interacting model we can 
compute the time evolution $E_0(t)$ of the ``internal'' energy
\bea
E_0(t)&\equiv&\langle \psi(t)|\hat H_0|\psi(t)\rangle,\label{E_t}\\
|\dot \psi(t)\rangle&=&-i \hat H(t)|\psi(t)\rangle,
\eea
with $|\psi(t=0)\rangle=|\psi_0\rangle$, i.e. we start in the GS of the non-hybridized Hamiltonian $H_0$.
For our purposes this quantity is more helpful than the ``total'' energy $E(t)\equiv\langle \psi(t)|\hat H(t)|\psi(t)\rangle$, which is stepwise constant,
because $E_0(t)$ is a continouos function of time, and shows a non trivial evolution when $\hat H(t)=\hat H_1$, whereas it is still obviously constant when $\hat H(t)=\hat H_0$.
However, both expressions $E_0(t)$ and $E(t)$ lead to the same dissipation per cycle (they are equal when $\hat H(t)=\hat H_0$), 
so as to provide the same outcome when predicting the results of experiments.
In what follows we study in detail $E_0(t)$ in different situations.

\section{Numerical results}\label{sec_num_res}
Let us examine 
the numerical results for $E_0(t)$.

\subsection{Single quench}
\begin{figure}[bt]
\includegraphics[width=0.45\textwidth]{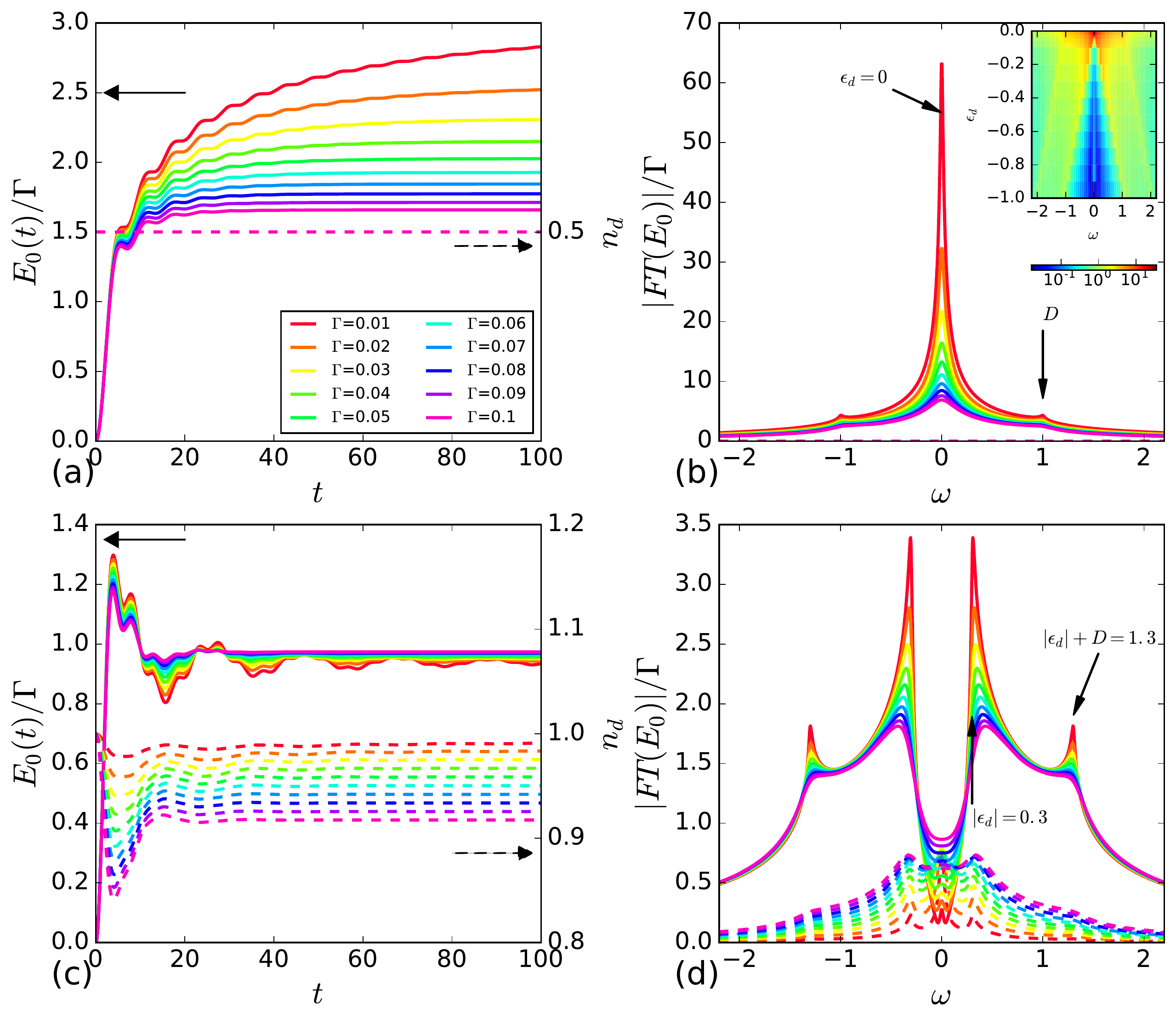}
\caption{(a) Internal energy $E_0(t)\equiv \langle \psi(t)| \hat H_0|\psi(t)\rangle$ as a function of time for a single quench at $t=0$ from $\hat H_0$ to $\hat H_1$
and occupation of the $d$ level $n_d(t)$ (dashed line), here identically equal to $0.5$.
We set at zero the initial energy $E_{00}\equiv \langle\psi_0| H_0|\psi_0\rangle$, where $|\psi_0\rangle$ is the GS of $H_0$;
we use $\epsilon_d=0$ and $\rho_{flat}$.
(b) Fourier transform (FT) $E(\omega)$ of $E(t)-E(t\rightarrow \infty)$, with peaks at zero energy and at $D$:
in the inset colormap of the FT for $\Gamma=0.01$ and different values of $\epsilon_d$, showing the evolution of the two peaks at $|\epsilon_d|$ and $|\epsilon_d|+D$.
(c)-(d) Same for $\epsilon_d=-0.3$; 
in the FT we can observe peaks at $|\epsilon_d|=0.3$ and $|\epsilon_d|+D=1.3$.
}\label{fig_static}
\end{figure}

Starting from a single quench, with Hamiltonian:
\bea
\hat H(t)=
\begin{cases}
\hat H_0, & t<0,\\
\hat H_1, & t\geq 0.
\end{cases}
\eea
we obtain results for $\rho_{flat}$ as reported in Fig. \ref{fig_static}.
The energy starts at $E_{00}\equiv \langle \psi_0|\hat H_0|\psi_0\rangle\equiv 0$ at $t=0$, jumps at each switch with 
damped oscillations, 
each time relaxing to a final value $E^0_{inf}=E_{00}+E_{diss}^{\tau\rightarrow\infty}$ in a time scale $\sim 1/\Gamma$. 
The parameter $\Gamma$ dictates how fast the system relaxes to its new steady-state situation;
$E_{diss}^{\tau\rightarrow\infty}$ is given by Eq. \eqref{e_diss_1_ex} up to $1/L$ and $\Gamma/D$ corrections.

We 
note 
that the system never relaxes to a time independent wave-function $|\psi_{inf}\rangle$, since the evolution is Hamiltonian,
but all the observables do (before a certain time $t^*(2L)$ for which we can observe quantum revivals, in our case $t^*(600)\simeq 1800$). 
From the Fourier transform $E_0(\omega)$ of $E_0(t)$ we can observe peaks at frequencies equal to $|\epsilon_d|$ and, to a minor extent, $|\epsilon_d|+D$;
these peaks cannot be easily fitted by Lorentzians, showing that all the frequencies in between matter, and the relaxation cannot be described by simple exponential terms.
Indeed, the initial rise of $E_0(t)$ is $\Gamma$-independent and dictated by the high-frequency term $|\epsilon_d|+D$;
only at larger time is the decrease exponential with decay rate $\Gamma$.
For the "Kondo" case $\epsilon_d=0$, we are left with weak oscillations at high frequency $\sim D$, and the approach to the final energy is (almost) monotonouos.
We also consider the time evoution of the occupation of the $d$-level:
\be\label{nd}
n_d(t)=\langle \psi(t)|\hat n_d|\psi(t)\rangle, \hspace{5pt} \hat n_d=\sum_\sigma \hat d_\sigma^\dagger \hat d_\sigma,
\ee
which is identically $0.5$ for $\epsilon_d=0$, since the system is always at particle-hole symmetry,
but has a non-trivial time evolution for $\epsilon_d=-0.3$, with peaks in the FT at the same values as $E_0(\omega)$.

\begin{figure}[bt]
\includegraphics[width=0.45\textwidth]{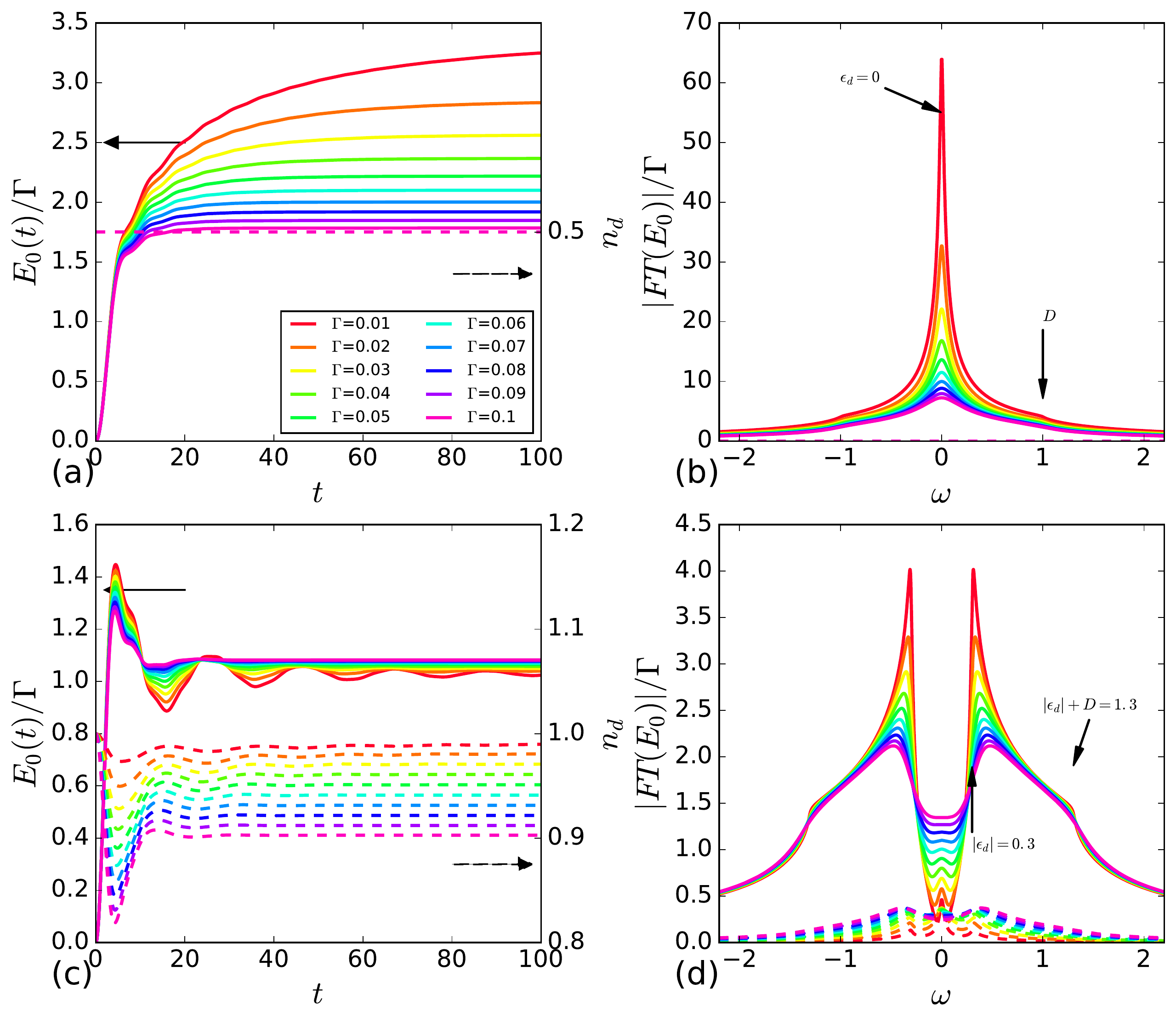}
\caption{Same as Fig. \ref{fig_static} but for a semicircular DOS $\rho_{sqrt}$.
Oscillations at $\omega=|\epsilon_d|+D$ are less pronounced, 
as the peak in the FT of Fig. \ref{fig_static} is here a shoulder.
}\label{fig_static_sqrt}
\end{figure}

Using $\rho_{sqrt}$, Fig. \ref{fig_static_sqrt}, we can observe that the peak at $|\epsilon_d|$ remains, so it is a robust feature, 
while the one at $|\epsilon_d|+D$ is almost completely washed away, so it is likely an artifact of $\rho_{flat}$.
One last important quantity is the time $\bar t$ at which the energy reaches its first local maximum: 
it is found to be independent from $\Gamma$, and roughly equal to $\bar t \sim \pi/(|\epsilon_d|+D)$.
We will see that this approximatively corresponds to the optimal half-period for which the dissipated power is maximal.

\subsection{Time-periodic driving}
We can now switch to time-periodic driving, with Hamiltonian:
\bea
\hat H(t)=
\begin{cases}
\hat H_0, & t<0 \mbox{ or } (2p+1)\tau <t< (2p+2)\tau,\\
\hat H_1, & 2p\tau <t< (2p+1)\tau,
\end{cases}
\eea
$p\geq 0$, and period $2\tau$.

\begin{figure}[bt]
\includegraphics[width=0.45\textwidth]{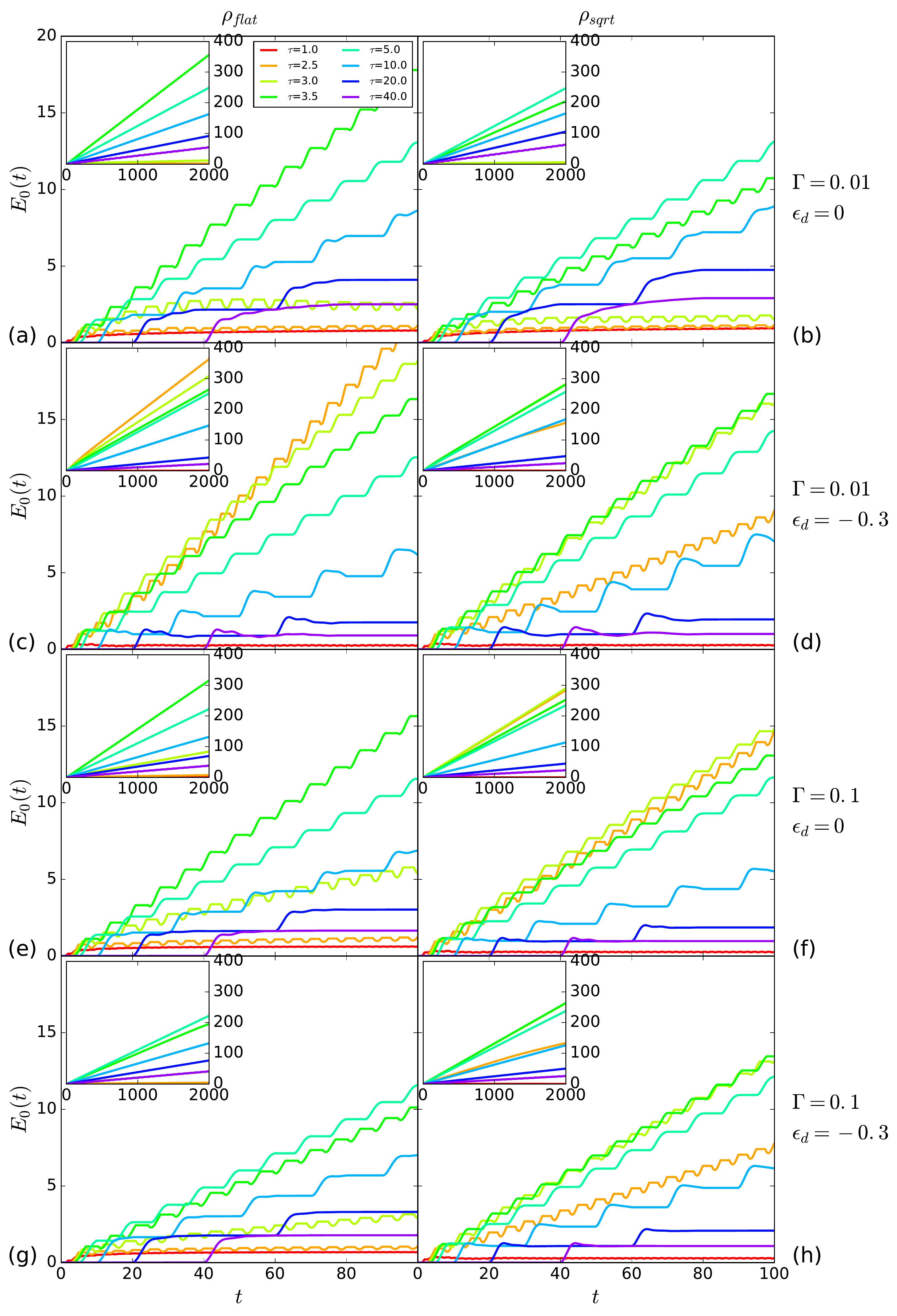}
\caption{Time evolution of the internal energy $E_0(t)$ for a periodic driving at different semiperiods $\tau$, with:
(a) $\Gamma=0.01$, $\epsilon_d=0$ and $\rho_{flat}$,
(b) $\Gamma=0.01$, $\epsilon_d=0$ and $\rho_{sqrt}$,
(c) $\Gamma=0.01$, $\epsilon_d=-0.3$ and $\rho_{flat}$,
(d) $\Gamma=0.01$, $\epsilon_d=-0.3$ and $\rho_{sqrt}$,
(e) $\Gamma=0.1$, $\epsilon_d=0$ and $\rho_{flat}$,
(f) $\Gamma=0.1$, $\epsilon_d=0$ and $\rho_{sqrt}$,
(g) $\Gamma=0.1$, $\epsilon_d=-0.3$ and $\rho_{flat}$,
(h) $\Gamma=0.1$, $\epsilon_d=-0.3$ and $\rho_{sqrt}$.
The system almost immediately reaches the steady-state, i.e. transient effects are weak (but not totally negligible).
In the inset we plot the evolution at large times, which clearly shows the linear increase of the energy at long times for large enough $\tau$.
}\label{fig_e_t}
\end{figure}

In Fig. \ref{fig_e_t}  we plot $E_0(t)$ for different values of $\tau$, using $\Gamma=0.01,0.1$, $\epsilon_d=0,-0.3$, and $\rho=\rho_{flat},\rho_{sqrt}$.
We can see that, at small $\tau$, the power is zero, because the energy goes to a constant.
When $\tau =\tau^* \sim \pi/(|\epsilon_d|+D)$, the energy starts to increase linearly per period,
so the power is a constant function of time.
At larger $\tau$, the energy increase per cycle saturates, and the power goes like $1/\tau$.

We can track this behavior by we plotting $\Delta E(\tau$) and $P(\omega)$ for different values of $\tau$.
Here it is enough to compute the energy at each plateau by:
\be
E[2p\tau]=\langle \psi_0| (e^{i\hat H_1\tau}e^{i\hat H_0\tau})^p \hat H_0 (e^{-i\hat H_0\tau}e^{-i\hat H_1\tau})^p|\psi_0\rangle,
\ee
and extract the energy increase per cycle and power as:
\bea
\Delta E(\tau)&=& E[2p\tau]-E[2(p-1)\tau],\\
P(\tau)&=&\frac{\Delta E(\tau)}{2\tau},\\
P(\omega)&=& P(\pi/\tau),
\eea
when $p \gg 1$, i.e. in the stationary state, where these quantities do not depend on $p$.

\begin{figure}[bt]
\includegraphics[width=0.45\textwidth]{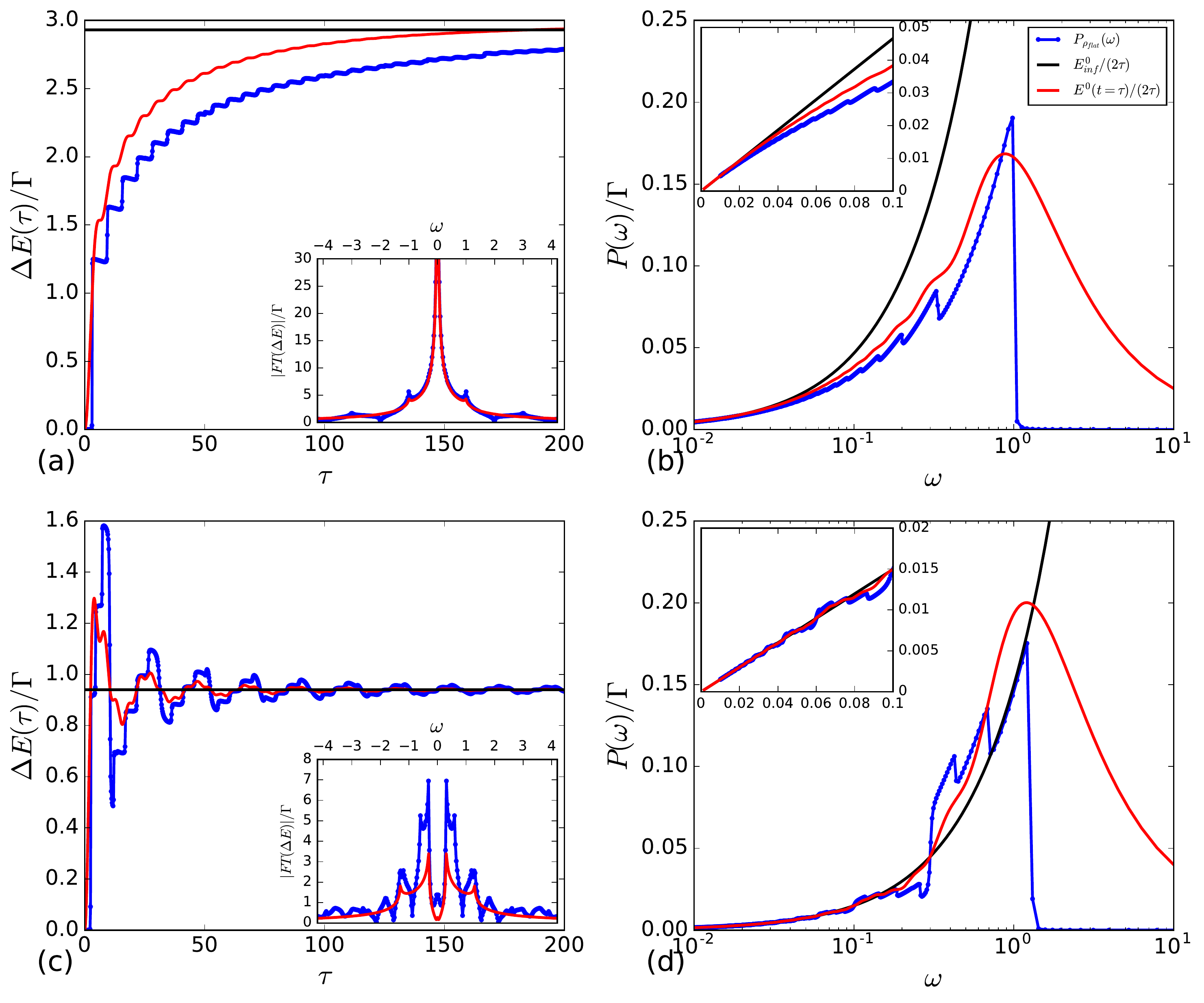}
\caption{(a) Energy dissipation per period $\Delta E(\tau)\equiv 2\tau P(\tau)$ in the steady-state as a function of $\tau$, for $\epsilon_d=0$, $\Gamma=0.01$.
We also plot the time evolution of the internal energy for a single quench $E_0(t=\tau)$, and its asintotic value $E_{diss}^{\tau\rightarrow\infty}$.
In the inset, the Fourier transform of $E_0(t=\tau)$ and $\Delta E(\tau)$.
(b) Power dissipation $P(\omega)$ (energy dissipation per unit of time) as a function of $\omega=\pi/\tau$: in the inset the linear behavior at small frequencies.
(c)-(d) Same for $\epsilon_d=-0.3$.
}\label{fig_p_t}
\end{figure}

\begin{figure}[bt]
\includegraphics[width=0.45\textwidth]{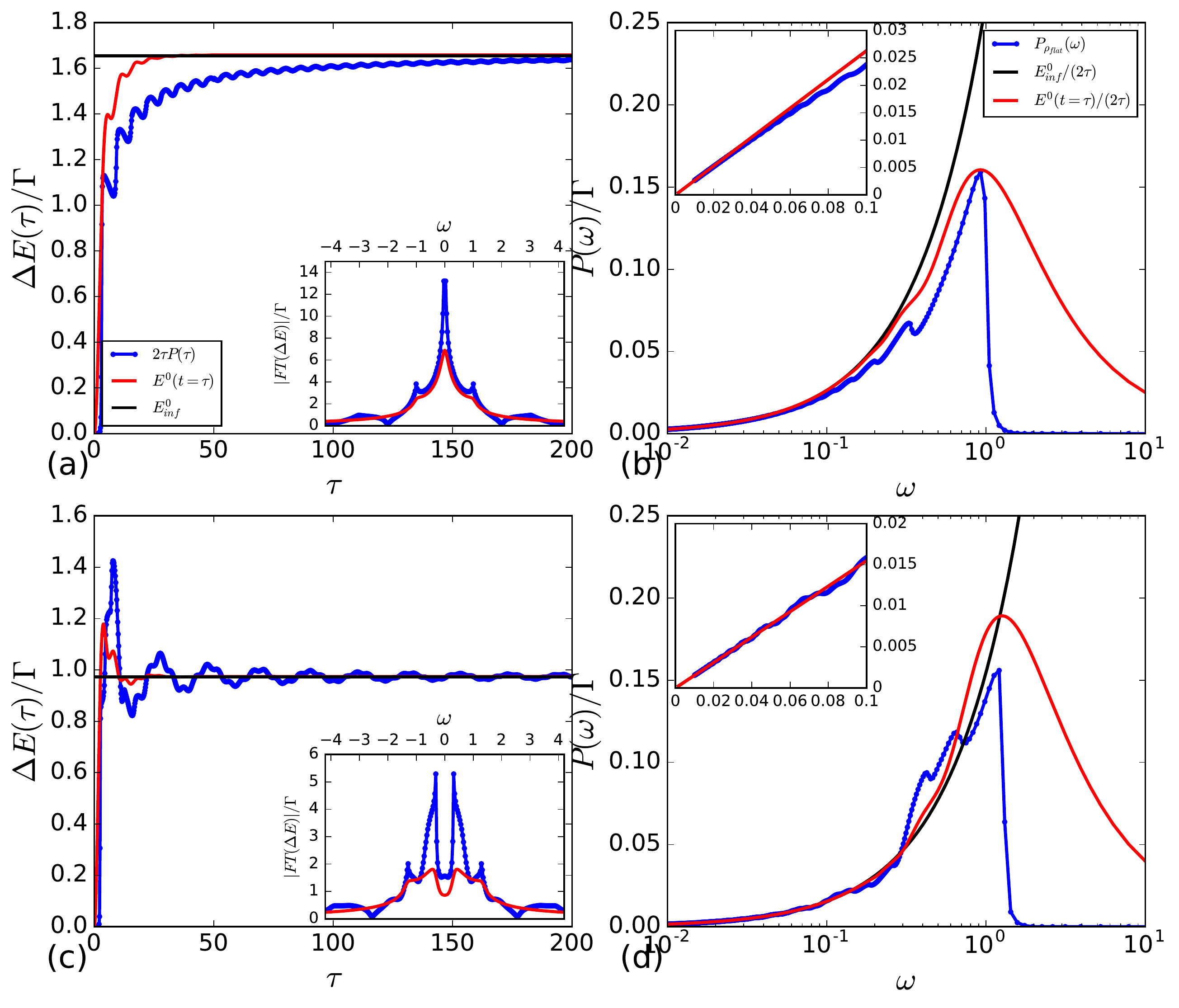}
\caption{Same as Fig. \ref{fig_p_t} but for $\Gamma=0.1$.
}\label{fig_p_t_01}
\end{figure}

\begin{figure}[bt]
\includegraphics[width=0.45\textwidth]{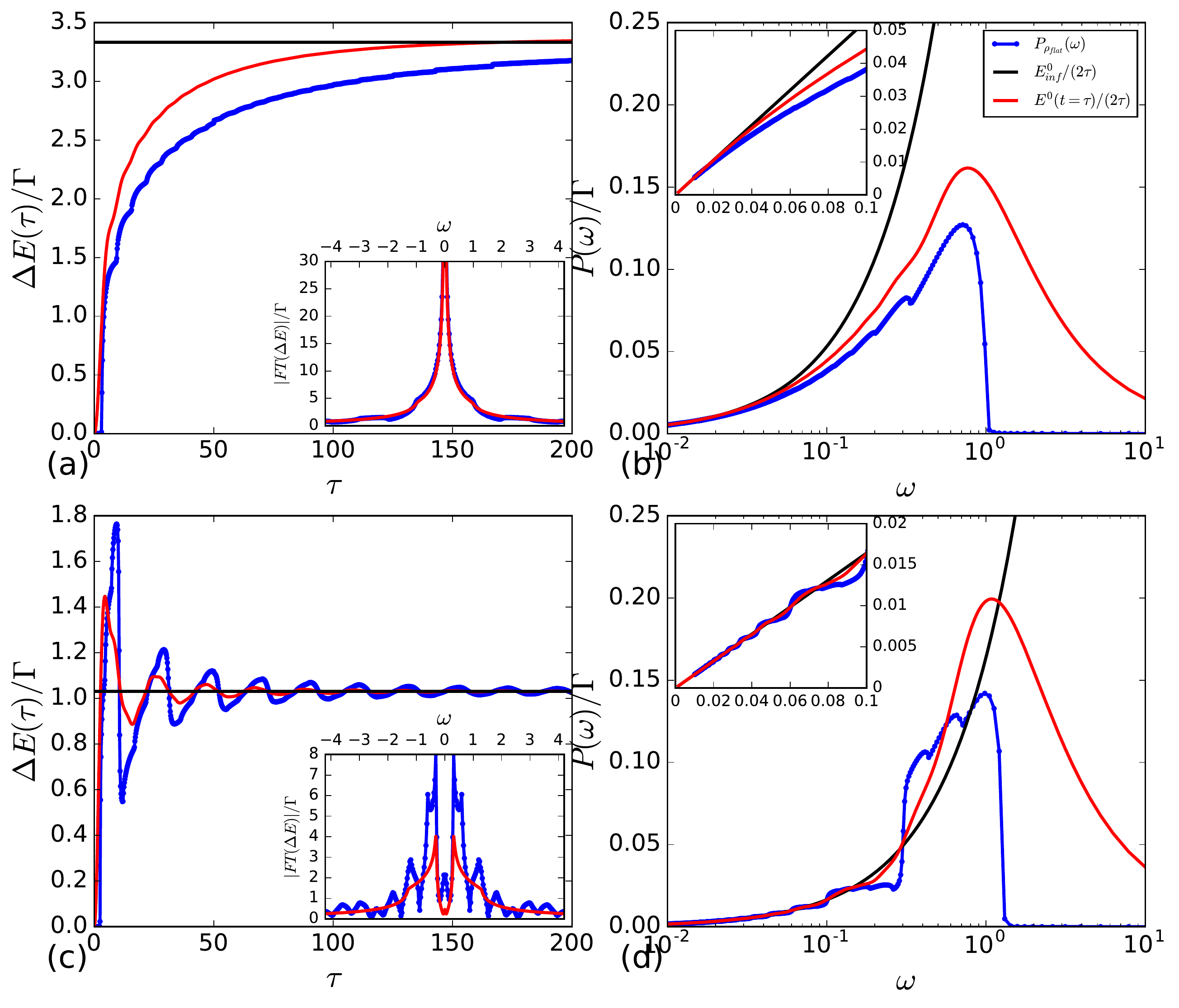}
\caption{Same as Fig. \ref{fig_p_t} but for $\rho_{sqrt}$.
}\label{fig_p_t_sqrt}
\end{figure}

\begin{figure}[bt]
\includegraphics[width=0.45\textwidth]{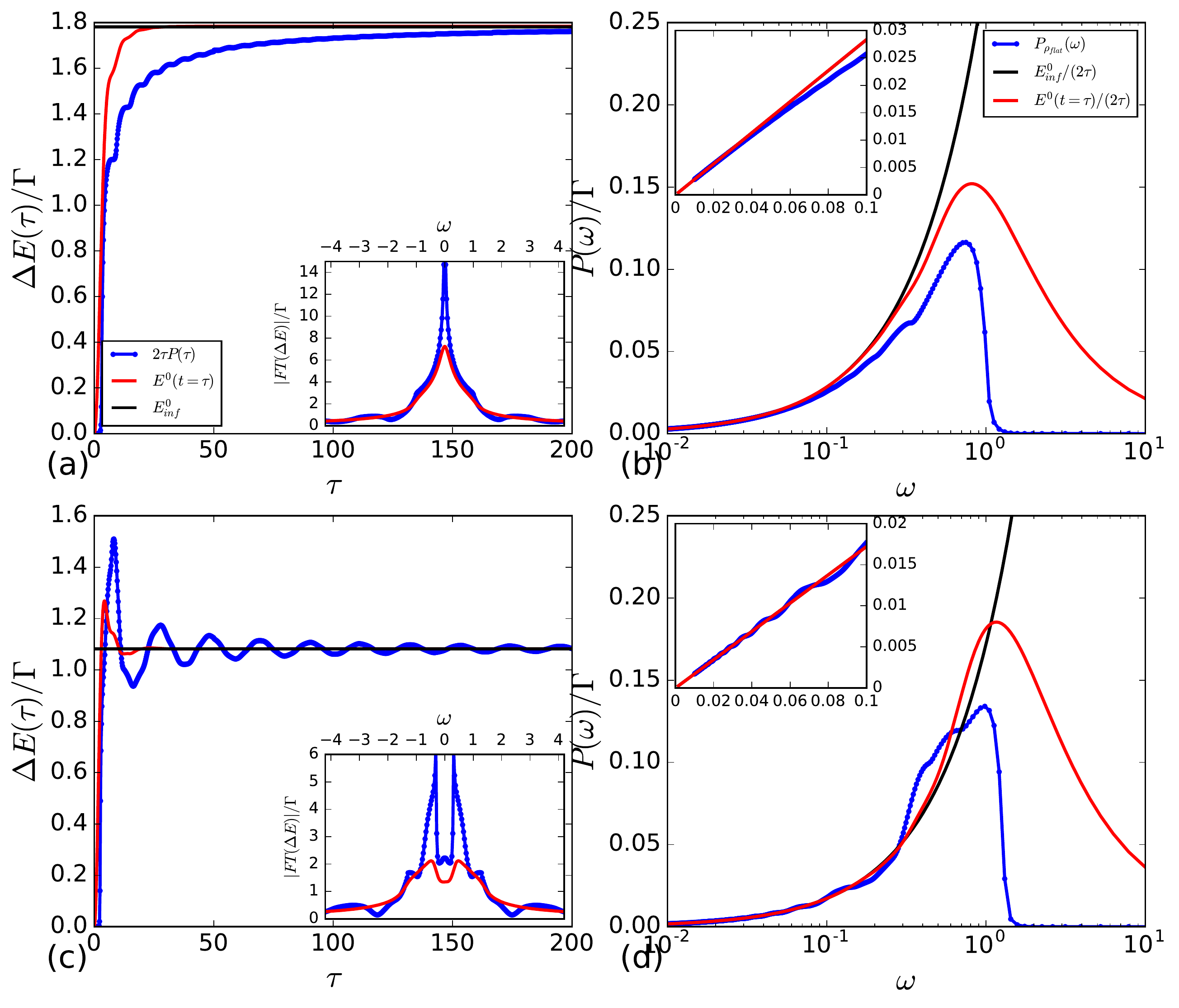}
\caption{Same as Fig. \ref{fig_p_t_01} but for $\rho_{sqrt}$.
}\label{fig_p_t_01_sqrt}
\end{figure}

We show results in Figs. \ref{fig_p_t}, \ref{fig_p_t_01}, \ref{fig_p_t_sqrt}, and \ref{fig_p_t_01_sqrt},
where we take either $\Gamma=0.01$ or $\Gamma=0.1$, and either a flat or a semicircular density of states (for a total of 4 cases).
We see that they all show the same qualitative behavior, so let's analyse their general properties.

%



\subsubsection{Low frequencies}
Low frequencies correspond to driving times $\tau \gg 1/\Gamma$, i.e. the system can reach its steady state 
before it undergoes a new quench: the internal energy $E_0(t)$ can reach a constant value during each semiperiod.

We can assume that, if after each cycle the system gains the same energy $E_0(t\gg 1/\Gamma)\simeq E_{diss}^{\tau\rightarrow\infty}$ 
that it gains after the first cycle, the power is simply given by:
\be\label{p_lowf}
\lim_{\tau\rightarrow \infty}P(\tau)\sim \frac{E_{diss}^{\tau\rightarrow\infty}}{2\tau}
\ee
so, at low frequencies, the power is linear in the frequency.
Numerics show that this 
assumption
is extremely good:
after a few cycles, the system absorbs a constant amount of energy per cycle which is very close to $E_{diss}^{\tau\rightarrow\infty}$.

\subsubsection{Intermediate frequencies}
For intermediate frequencies we can 
similarly note that
after the first quench, $E_0(t)$ evolves as for a single quench, and, when the Hamiltonian goes back to $H_0$,
it attains a plateau at energy $E_0(t=\tau)$.
Now, if during the half-period in which $H_0$ acts, the systems ``forgets'' completely about its history, 
it is in an effective GS but with an energy equal to $E_{00}+E_0(\tau)$:
so, when we perform a new quench to $H_1$, at the end of the new cycle it will gain another $E_0(\tau)$,
for a total of $E_{00}+2E_0(\tau)$, and so on.
Henceforth, in this approximation, the energy gain per cycle as a function of $\tau$ is equal to the time evolution at $E^0(t=\tau)$ for a single quench:
\be
P(\tau)\sim \frac{E_0(t=\tau)}{2\tau}.
\ee
Any difference betweenbetween this simple expression and the exact numerical result is related to the ``memory'' that the system has of its previous quenches, and, in particular, of what happens during the initial transient.
This approach is valid down to low frequencies, when $E_0(\tau)$ converges to $E_{diss}^{\tau\rightarrow\infty}$, leading to Eq. \eqref{p_lowf}.

We can see in the numerical results that $\Delta E(\tau)$ and $E_0(t=\tau)$ are in general very similar, in particular for large $\tau$, and oscillate at the same frequencies, 
even though $E_0(t=\tau)$ has more pronounced oscillations.

\subsubsection{High Frequencies}
The limit of high frequencies corresponds to driving times $\tau \sim  1/(|\epsilon_d|+D)$,
i.e. frequencies on the order of $|\epsilon_d|$, or of $D$ if $\epsilon_d=0$.
The maximum
dissipated power is attained around this frequencies;
this is due to the fact that $E_0(t=\tau)$ grows very fast for small $\tau$,
independently of $\Gamma$, reaching a considerable fraction of $E_{diss}^{\tau\rightarrow\infty}$
already at $\tau \sim 1/D$:
\bea
\max P(\omega)\equiv P(\omega^*)&\sim& \omega^* E_{diss}^{\tau\rightarrow\infty},\\
\omega^*&\sim& |\epsilon_d|+D.
\eea

\subsubsection{Ultra-High Frequencies}
At 
extremely 
high driving frequencies $\tau \ll 1/(|\epsilon_d|+D)$,
i.e. 
larger than all typical energy scales of the system,
we can apply Trotter's formula
:
\bea\label{eq_trotter}
&&e^{i\hat H_1\tau}e^{i\hat H_0\tau}\simeq e^{i(\hat H_1+\hat H_0)\tau}=\nonumber\\
&&=e^{i[(\hat H_1+\hat H_0)/2](2\tau)}=e^{i\hat H_{eff}(2\tau)}
\eea
adequate because 
the system evolves as if the Hamiltonian were $\hat H_{eff}=(\hat H_1+\hat H_0)/2$ for the whole cycle.
Now $(\hat H_1+\hat H_0)/2=\hat H_1[V/2]$, which corresponds to a hybridized Hamiltonian with an effective hybridization $V_{eff}=V/2$ one half of the original one,
and an effective broadening $\Gamma_{eff}=\Gamma/4$ one fourth of the original one. 
Hence, in this limit the system converges to a steady state in which the energy is constant,
As a consequence, after the initial transient the power is zero:
\be
\lim_{\tau\rightarrow 0}P(\tau)= 0.
\ee
The system cannot respond to frequencies larger than the bandwidth, 
therefore cannot absorb any energy and the dissipation goes to zero.
This is an example of ``quantum Zeno effect'' \cite{quantum_zeno}.
Indeed, we see from the numerics that $P(\omega)$, after reaching its maximum, drops extremely fast to zero.

\begin{figure}[bt]
\includegraphics[width=0.45\textwidth]{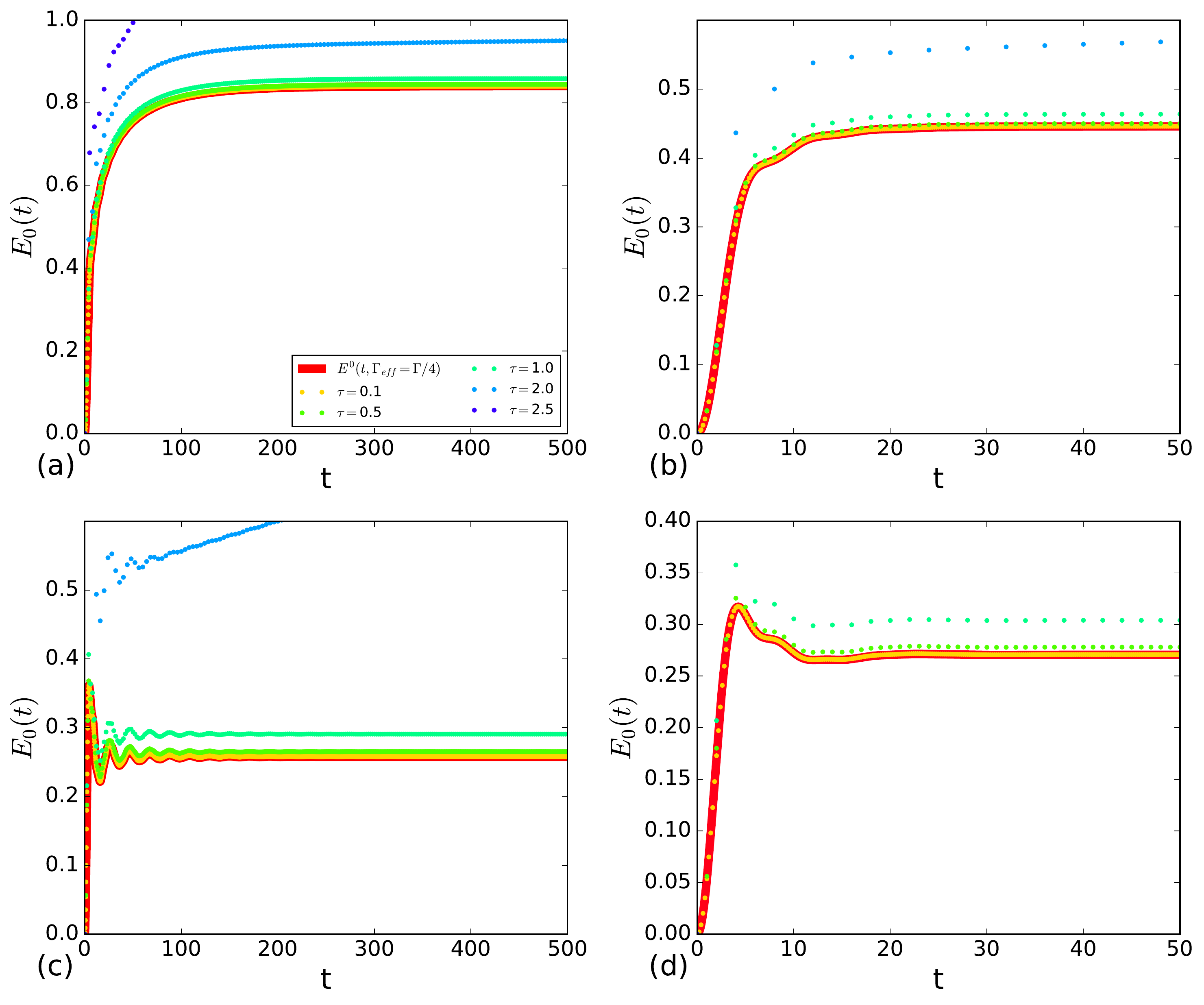}
\caption{Time evolution of $E_0(t)$ in the high-frequency limit (taken at each plateau), compared with its evolution for a single quench at $\Gamma/4$, with $\rho=\rho_{sqrt}$:
(a) $\Gamma=0.01$, $\epsilon_d=0$
(b) $\Gamma=0.1$, $\epsilon_d=0$
(c) $\Gamma=0.01$, $\epsilon_d=-0.3$
(d) $\Gamma=0.1$, $\epsilon_d=-0.3$.
}\label{fig_trotter}
\end{figure}

In Fig. \ref{fig_trotter}, we 
address the high-frequency limit $\omega>\omega^*$ in more detail.
By plotting
$E(t)$ for very small $\tau$, and comparing  it to $E_0(t=\tau)$
for a system with $\Gamma_{eff}=\Gamma/4$, as suggested by the Trotter's formula, Eq. \eqref{eq_trotter}, we see that the agreeement is very good for $\tau \ll 1/(|\epsilon_d|+D)$.



\subsection{Anderson impurity}
In the previous Sections we presented results for non-interacting impurities with different
values of the on-site energy $\epsilon_d$ and of the hybridization $\Gamma$.

However,  actual Kondo effect are realized 
by Anderson impurities (SIAM), 
which cannot be treated using the results and methods above. 

We 
found
in Ref. \cite{baruselli2017} that,
when computing the energy dissipation in the static limit from Eq. \eqref{e_diss_1v1} for a SIAM using NRG, the
results could be easily interpreted in terms of the sum of contributions from 
the Hubbard bands, with energy $\pm\epsilon_d$ and hybridization $\Gamma$,
and from the Kondo peak, with zero energy and hybridization $T_K$,
where $T_K\ll \Gamma$ is the Kondo energy scale.

In order to study the dynamics, as we now wish, 
we should focus on what would happen if we 
were to follow $E_0(t)$ for a SIAM.
For a single quench, the contribution from Hubbard bands would quickly converge in a timescale $\sim 1/\Gamma$.
Then, the Kondo effect would buildup in a timescale $\sim 1/T_K\gg 1/\Gamma$.
When looking at the frequency-dependent dissipation, we would thus observe the sum of the two signals from 
the Hubbard bands and the Kondo resonance.

\begin{figure}[bt]
\includegraphics[width=0.45\textwidth]{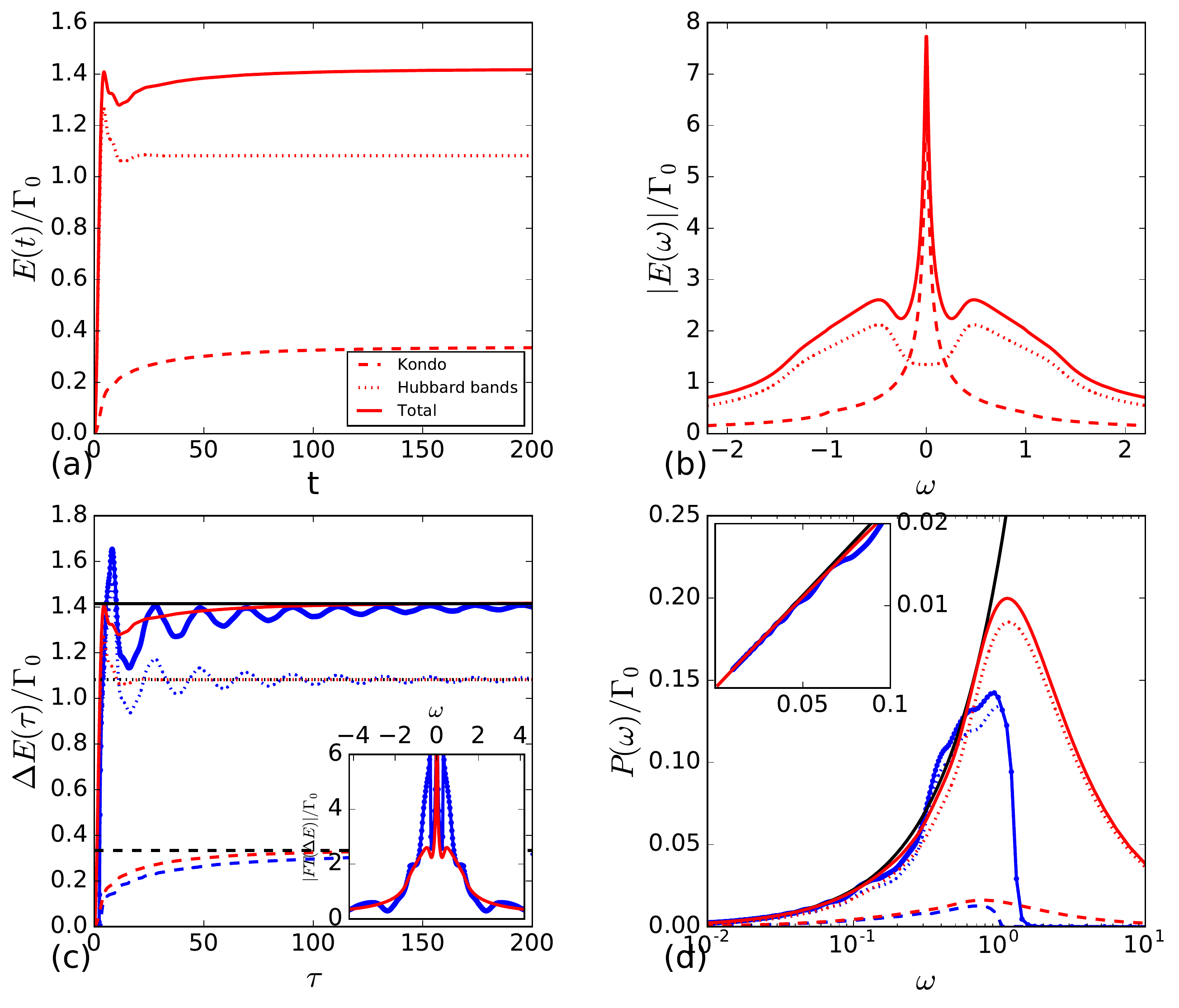}
\caption{Naive approach to an Anderson impurity:
we plot the sum of the dissipaed energy from a Kondo-like peak ($\epsilon_{d1}=0$, $\Gamma_1=0.01$)
and from Hubbard bands ($-\epsilon_{d0}=\epsilon_{d0}+U=0.3$, $\Gamma_0=0.1$); we use $\rho_{sqrt}$.
(a) Single quench and (b) its FT.
(c) Periodic switching. 
(d) Power.
}\label{fig_anderson}
\end{figure}

Of course, the two processes will generally interfere, but this goes unfortunately  beyond our one-body description.
We report in Fig. \ref{fig_anderson} the crude result 
in which we sum the contributions from the Hubbard bands and the Kondo peak.

The total dissipation in the low frequency limit would be proportional to $\Gamma+\Gamma_K\sim \Gamma$;
however, in the frequency dependence we could observe different regimes, with peaks around $|\epsilon_d|$ and broadening $\sim \Gamma$ for the Hubbard bands,
and peak around zero and broadening $\sim T_K$ for the Kondo resonance.

This simplified approach of course neglects many details,
in particular the spin dynamics, but it can 
hopefully serve as a 
useful starting point for more detailed calculations.


\subsection{Changing the hybridization}
\begin{figure}[bt]
\includegraphics[width=0.45\textwidth]{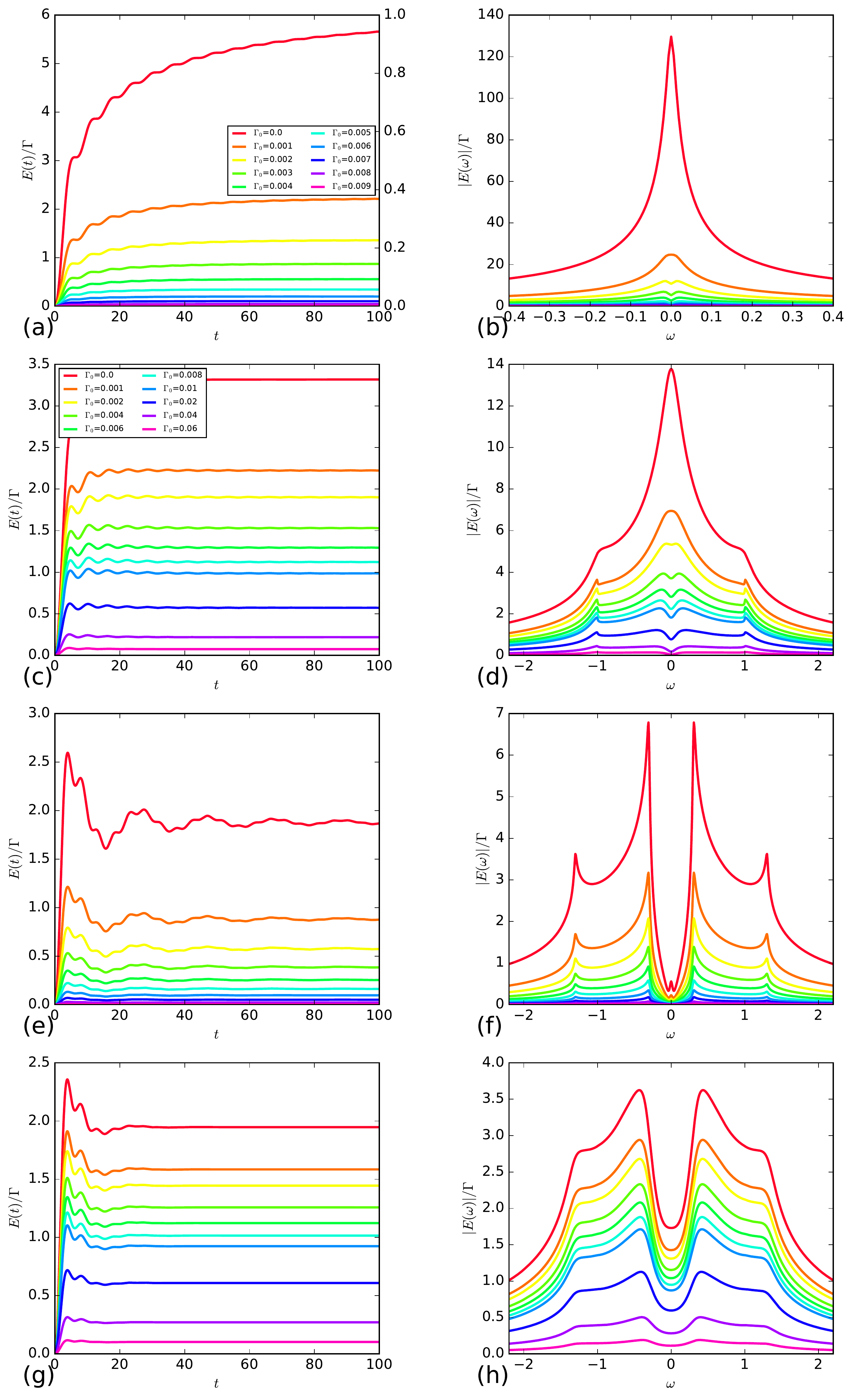}
\caption{On the left time evolution of $E_0(t)$ for a single quench $\Gamma_0\rightarrow\Gamma_1$.
(a-b): $\Gamma_1=0.01$, $\rho=\rho_{flat}$,
(c-d): $\Gamma_1=0.1$, $\rho=\rho_{flat}$,
(e-f): $\Gamma_1=0.01$, $\rho=\rho_{sqrt}$,
(g-h): $\Gamma_1=0.1$, $\rho=\rho_{sqrt}$.
}\label{fig_V1V2}
\end{figure}

Up to now we have considered quenches where we completely switch the hybridization off.
However, it is also possbile to consider quenches $\Gamma_1 \rightarrow \Gamma_2$,
where both $\Gamma_1$ and $\Gamma_2$ are positive.
As we are going to see, we must follow this approach in the TD-Gutzwiller approach.
The Hamiltonian is here:
\bea
\hat H_0&=&\hat T +\hat H_d +V_0\hat v,\\
\hat H_1&=&\hat T +\hat H_d +V_1\hat v,\\
\hat V&=&(V_1-V_0)\hat v.
\eea
It can be shown that at zero frequency, when $\epsilon_d=0$, the energy dissipation is:
\be\label{diss_v1v2}
E_{diss}^{\tau\rightarrow\infty}=\frac{4}{\pi}(V_1-V_0) \left( \frac{\Gamma_0}{V_0} \log \Gamma_0 - \frac{\Gamma_1}{V_1} \log \Gamma_1\right).
\ee
We present in Figs. \ref{fig_V1V2} results for this case.
It can be seen that the total dissipation decreases, as expected, with respect to the case $\Gamma_1=0$.

\section{Gutzwiller approximation}\label{section_ga}
Let's now switch to the study of a SIAM.
We will use a mean-field approach, namely the Gutzwiller approximation.

\subsection{Time-independent theory}
Given a SIAM:
\be
\hat H=\hat T+\hat V_h + \hat U,
\ee
with 
\be
\hat U=\frac{U}{2}(\hat n_d-1)^2,
\ee
$\hat n_d$ from Eq. \eqref{nd}, $\hat T$ from Eq. \eqref{eq_T}, and $\hat V_h$ from Eq. \eqref{eq_v},
the Gutzwiller approximation consists in minimizing the variational energy
$E_{tot}=\langle \phi| \hat H |\phi \rangle$
where $|\phi\rangle = \mathcal{\hat P}_d |\psi\rangle$, $|\psi\rangle$ is a Slater determinant, GS of a RLM,
and  $\mathcal{\hat P}_d$ is a local projector on the $d$ states:
\bea
\mathcal{\hat P}_d&=&\sqrt{2P} \mathcal{\hat P}_{02}+\sqrt{2-2P} \mathcal{\hat P}_{1},\\
\mathcal{\hat P}_{02}&=&|\uparrow\downarrow\rangle\langle\uparrow\downarrow|+|0\rangle\langle0|,\\
\mathcal{\hat P}_{1}&=&|\uparrow\rangle\langle \uparrow|+|\downarrow\rangle\langle \downarrow|.
\eea
Here $P/2$ is the probability of double occupation on the impurity:
when $U=0$, this double occupation is $1/4$, so $P=1/2$;
In general, $0\leq P \leq 1$, but, for a repulsive $U$, $0\leq P \leq 1/2$.

Minimization of the energy expectation value for $\rho_{flat}$:
\bea
E_{tot}&=&\langle\psi|\hat H_R|\psi\rangle+\frac{UP}{2}=\\
&=&\frac{2\Gamma z}{\pi}\log\frac{z\Gamma}{e}+\frac{UP}{2},
\eea
with 
\bea
z&=&4P(1-P),\\
\hat  H_R&=&\hat  T+R\hat V_h,\\
R&=&\sqrt{z},
\eea
leads to the SCF equation:
\be\label{scf_z}
\sqrt{1-z}\log \Gamma z=-\frac{\pi U}{16 \Gamma}.
\ee
We define the (Gutzwiller approximation of the) Kondo temperature $T_K$ as $z\Gamma$.
For $U=0$, $T_K=\Gamma$; 
for large $U/\Gamma$, we find:
\be
T_K\equiv z\Gamma \sim D \exp{\left( -\frac{\pi U}{16\Gamma}\right)}.
\ee
This exponential decrease of $T_K$ for large $U$ cannot be reproduced at an arbitrary precision when performing numerics.
When $T_K \sim \delta$, where $\delta$ is the level spacing for a finite system, we find that the SCF solution for $z$ drops suddenly to zero, as in the $\Gamma=0$ case for an infinite system.
Hence, we will not work in the regime $U \gg \Gamma$, but keep $U\sim \Gamma$.

The energy $E$ and the expectation value of the hybridization $\mathcal{V}$ for an infinite system are:
\bea
E=\langle \hat H_R \rangle &=& \frac{2\Gamma z}{\pi}\log\frac{z\Gamma}{e},\label{e_tot}\\
\mathcal{V}=\langle \hat V_h \rangle &=& \frac{4\Gamma \sqrt{z}}{\pi}\log z\Gamma.\label{v}
\eea

\subsection{Time-dependent theory}

We now consider the TD theory.
In this case $|\phi\rangle=\exp{(-i \theta  \mathcal{\hat P}_{02})} \mathcal{\hat P}_d |\psi\rangle$,
$P(t)$ depend on time, and a new variational parameter $\theta(t)$ is introduced.
The equations of motion are:
\bea
\dot P&=&-\mathcal{V}\frac{\partial R}{\partial \theta},\\
\dot \theta&=&\frac{U}{2}+\mathcal{V}\frac{\partial R}{\partial P}, \\
\mathcal{V}&=&\langle \psi|\hat V_h|\psi\rangle,\\
|\dot\psi\rangle&=&-i \hat H_R |\psi\rangle,\\
\hat H_R&=&\hat T+R\hat V_h,\\
R&=&2\sqrt{P(1-P)}\cos\theta.
\eea

All these quantities are to be considered dependent on time:
$P(t)$, $\theta(t)$, $R(t)$, $\mathcal{V}(t)$, $|\psi(t)\rangle$, $\hat H_R(t)$.
The kinetic energy operator $\hat T$ is independent of time.
The hybridization $V(t)$ and the Hubbard repulsion $U(t)$ are the quenching variables, which we can control externally,
and are taken to be stepwise constant.

We will consider two kinds of quench: the interaction quench, in which $V=V_0=V_1$ is constant, and we modulate $U$ between $U_0$ and $U_1$;
and the hybridization quench, in which $U=U_0=U_1$ is constant, and we modulate $V$ between $V_0$ and $V_1$.

It is easy to show that the total energy
\be
E_{tot}=\langle \psi|\hat H_R|\psi\rangle+PU/2,\\
\ee
is a constant of motion, as long as $U$ and $V$ do not depend explicitly on time;
it has a step-like behaviour when a quench in $U$ or $V$ is performed.

The expectation value $E_0\equiv \langle \hat H_R^0\rangle$ of 
\be
\hat H_R^0=\hat T+RV_0\hat v +PU_0/2
\ee
coincides during the half period in which $V=V_0$ and $U=U_0$ with $E_{tot}$, hence it is constant, and it evolves non-trivially during the other half-period.
The opposite happens for 
the expectation value $E_1\equiv \langle \hat H_R^1\rangle$ of 
\be
\hat H_R^1=\hat T+RV_1\hat v +PU_1/2.
\ee
Both $E_0$ and $E_1$ are continuous functions of time, i.e. do not have jumps at a quench, in contrast with $E_{tot}$.

\section{Quenches in TD-Gutzwiller}\label{ga_quench}
Let's consider some general features of the time evolution in the TD-Gutzwiller approximation after a quench in either the Hubbard repulsion $U$
or the hybridization $V$.

\subsection{Interaction quench}
Let's start with modulating $U$, which is the usual approach taken in the literature.

\subsubsection{$U_0=0$}
This is the standard case: at $t=0$ one turns the interaction on, and lets the system evolve; the hybridization is identically $\Gamma$.
$P$ starts from $P_0=1/2$, then relaxes to $P_1 < 1/2$, which is the SCF solution for $\Gamma$ and $U=U_1$;
similarly, $R$ starts from $1$, and relaxes to $2\sqrt{P_1(1-P_1)}<1$.

\subsubsection{$U_1=0$}
This case is trickier.
The system starts with $0<P_0<1/2$.
Setting $x=1-2P$, one gets:
\be\label{ddotx}
\ddot{x} +4 \mathcal{V}^2 x + \dot{\mathcal{V}} \dot x / \mathcal{V} + \mathcal{V} U_1 R=0,
\ee
which leads to Eq. \eqref{scf_z} when $\dot{x}=0$,
together with
\be
\dot R=-\frac{U_1}{2\mathcal V}\dot P.
\ee
So, when $U_1=0$, $R(t)=R_0$, and $\hat H_R$ is time-independent.
If $|\psi(t=0)\rangle$ is eigenstate of $\hat H_R$, then $\dot{\mathcal{V}}\propto \langle \psi(t)| [\hat H_R,\hat V_h]|\psi(t)\rangle=0$,
so Eq. \eqref{ddotx} becomes:
\be\label{ddotx_2}
\ddot{x} +4 \mathcal{V}^2 x =0,
\ee
with solution
\bea
x(t)=x_0 \cos\omega t,\\
\omega=-2 \mathcal V=-\frac{8\sqrt{z}\Gamma}{\pi} \log \Gamma z,
\eea
or 
\be
1-2P(t)=(1-2P_0) \cos\omega t.
\ee
If $|\psi(t=0)\rangle$ is not an eigenstate of $\hat H_R$, numerics show that $R(t)$ undergoes a transient, then converges to a finite value,
after which $P$ oscillates with the new constant value of $-2 \mathcal V$.

This shows that, when $U_1=0$, the system has a non-trivial dynamics, but it keeps oscillating and never relaxes to what would be the SCF solution $P=1/2$.
Hence, a quench to $U_1=0$ is a pathological case for the TD Gutzwiller method.

\subsection{Hybridization quench}
Following the first part of the paper, and Ref. \cite{baruselli2017}, we are going to study a hybridization quench,
in which we suddenly change the hybridization from $V_0$ (leading to $\Gamma_0=\pi\rho V_0^2$) to $V_1$ (leading to $\Gamma_1=\pi\rho V_1^2$),
leaving $U$ constant.

Let's consider some limiting cases.

\subsubsection{$U=0$}
Since the Gutzwiller approximation is exact when $U=0$, results for this case coincide with the non-interacting hybridization quench of the first part of the article;
in this case $P=1/2$ and $R=z=1$ identically.

\subsubsection{$V_0=0$}
In this case, $P(0)=R(0)=\mathcal{V}(0)=0$. 
This implies $\dot P=0$, henceforth $P(t)=R(t)=0$, which means that the system does not evolve.
The uncoupled solution $R=0$ is always a valid SCF solution, but an unstable one, i.e. it is a maximum of the energy, when $V\neq 0$.
However, this means that we cannot perform a quench where $V_0=0$.

\subsubsection{$V_1=0$}
The situation is similar when $V_1=0$.
In this case $P(0),R(0) \ne 0$, but $\mathcal{V}(t>0)=0$, so, once again, $\dot P=0$, and $P(t)=P_0$.
This means that we cannot perform a quench where $V_1=0$.

\subsubsection{General case}
The previous two results mean that, if we want to modulate periodically $V$ between $V_0$ and $V_1$, both of them must be nonzero.
In addition, since, if $U \gg \Gamma$, the numerical SCF solution gives $P=0$,
we must choose $V_1$ and $V_2$ large enough that the numerical SCF solution gives $P>0$ for both of them.
This add a complication in the analysis, since we have three parameters $U$, $V_0$, $V_1$ to consider.
In this case, $P$ starts from $0<P_0<1/2$, SCF solution for $U$ and $V_0$, and then relaxes to $0<P_1<1/2$, SCF solution for $U$ and $V_1$.

\section{Energy dissipation in TD-Gutwiller}\label{ga_energy}
Let's now consider the energy dissipation associated to a quench in the TD-Gutzwiller approximation.

\subsection{Zero frequency}
For the non-interacting case, in Ref. \cite{baruselli2017} we found that:
\be
E_{diss}^{\tau\rightarrow\infty}=\langle \psi_0| \hat V|\psi_0\rangle - \langle \psi_1| \hat V|\psi_1\rangle,
\ee
where $\hat V$ is the perturbation, $|\psi_0\rangle$ is the GS of $\hat H_0$, and $|\psi_1\rangle$ of $\hat H_1\equiv \hat H_0+\hat V$.
It is immediatley seen that this expression carries over to our SCF case, where in $\langle \psi_0| \hat V|\psi_0\rangle$ we must use the SCF parameters at $t=0$,
i.e. $P_0, R_0$, while for $\langle \psi_1| \hat V|\psi_1\rangle$ we need the SCF parameters at $t=\infty$, i.e. $P_1, R_1$.

This of course requires that the system thermalizes, which we have seen does not always happen, but we will exclude these pathological cases ($V_0=0$, $V_1=0$, or $U_1=0$).

\subsubsection{Interaction quench}
When we switch $U$ from $U_0$ to $U_1$ we get:
\be
E_{diss}^{\tau\rightarrow\infty}=\frac{(P_0-P_1)(U_1-U_0)}{2}.
\ee
We see that the dissipation is proportional to the Hubbard repulsion, i.e. it is much larger than the Kondo energy,
and it is not really useful for our purposes, so we will not treat this case anymore.

\subsubsection{Hybridization quench}
In this case we get:
\be
E_{diss}^{\tau\rightarrow\infty}=R_0\langle \psi_0| \hat V|\psi_0\rangle - R_1 \langle \psi_1| \hat V|\psi_1\rangle,
\ee
where $\hat V=(V_1-V_0)\hat v$. 
For an infinite system, using Eq. \eqref{v}, we get:
\be
E_{diss}^{\tau\rightarrow\infty}=\frac{4}{\pi}(V_1-V_0) \left( \frac{T_K^0}{V_0} \log T_K^0 - \frac{T_K^1}{V_1} \log T_K^1\right),
\ee
with $T_K^{0,1}=z_{0,1}\Gamma_{0,1}$.
In the limit $V_0\rightarrow 0$ this becomes:
\be
E_{diss}^{\tau\rightarrow\infty}=-\frac{4}{\pi}T_K^1\log T_K^1,
\ee
which is the energy dissipation for a quench of a Lorentzian level at zero energy and hybridization $T_K^1$;
see Eq. \eqref{e_diss_1_ex} with $\epsilon_d=0$.
Hence, in the $\tau \rightarrow \infty$ limit, the system is equivalent to a non-interacting level sitting at the Fermi energy
with hybridization given by the Kondo temperature. 
We will however see that the the behaviour at finite frequencies is different from that of a simple Lorentzian level.

\subsection{Finite frequencies}
We are interested in the hybridization quench with $U=U_0=U_1>0$ constant, and hybridization switched between $\Gamma_0>0$ and $\Gamma_1>0$.
In this case we must perform numerics.

\section{Numerical results for TD-Gutzwiller}\label{section_ga_num}
In this Section we present numerical results for the time evolution of a hybridization quench in the TD-Gutzwiller approximation.
We use $2L=100$ conduction states.

\subsection{Single quench}
As for the non-interacting case, we can we can gather a lot of information by looking at $E_0(t)$ for a single quench.
We recall that in that situation, oscillations at $|\epsilon_d|$ could be observed (so no oscillations for a Kondo-like impurity $|\epsilon_d|=0$),
and the asymptotic value was reached in a timescale $\sim 1/\Gamma$.

We also recall that here, when $U_1=0$, the system keeps oscillating at a frequence given by $\omega=-2\mathcal{V}\propto \Gamma$, and never relaxes.
When we increase $U$, we notice that the oscillating frequence slightly increases 
(we can approximatively fit it with $\omega\sim -2\mathcal{V}_1/\sqrt{z_1}\sim \Gamma_1 \log \Gamma_1 z_1$), while the relaxation rate increases steadily.

We come to the conclusion that, in the TD-dependent Gutzwiller approximation, the oscillation frequency is mainly set by $\Gamma$, 
and the relaxation rate by $U$.
This is quite the opposite of what happens in the non-interacting case.

\begin{figure}[bt]
\includegraphics[width=0.45\textwidth]{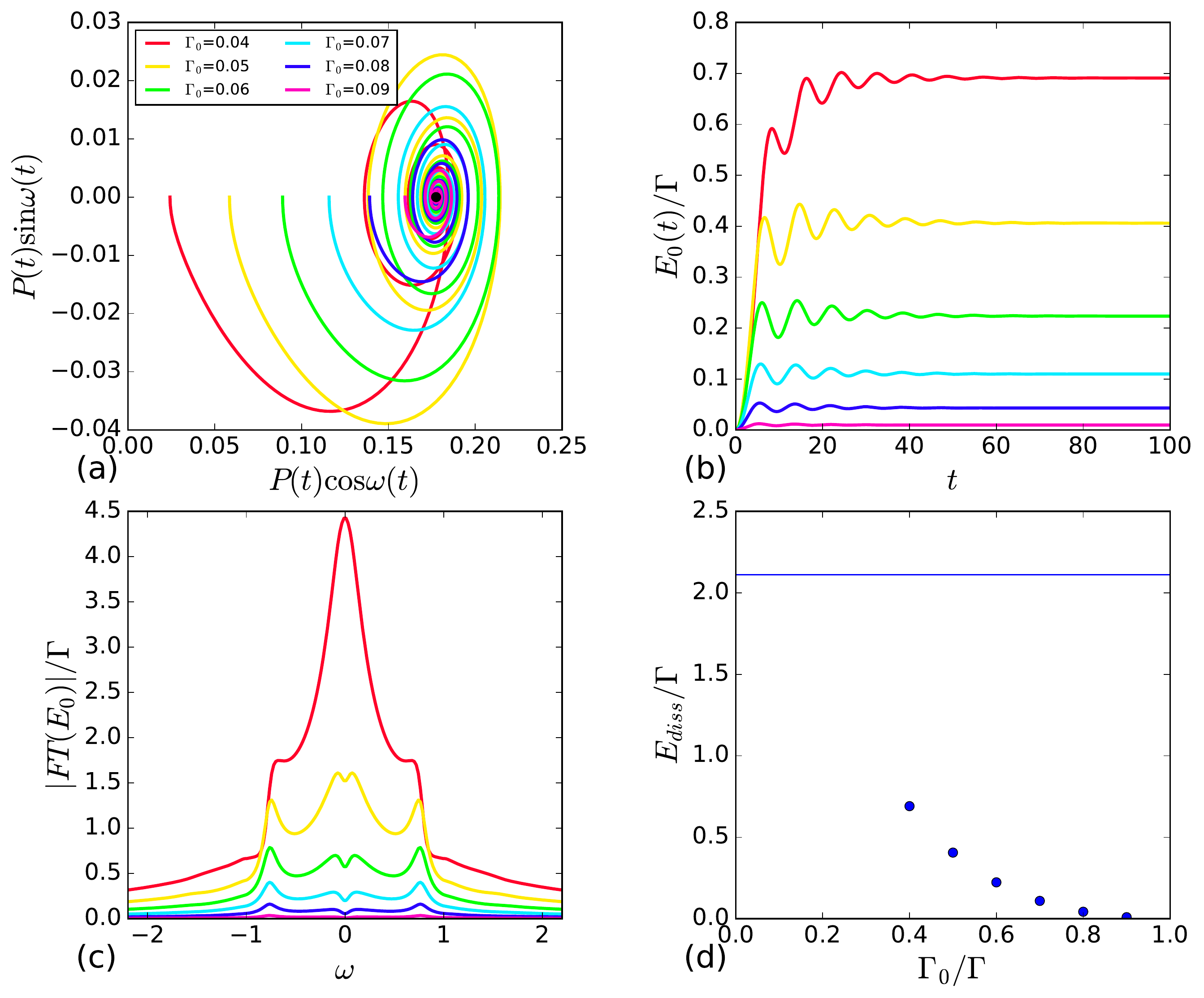}
\caption{Hybridization quench in the TD-Gutzwiller model for $\Gamma_1=0.1$, $U=1$ and different values of $\Gamma_0<\Gamma_1$.
(a) Time evolution of $P(t)$ and $\theta(t)$: all cases relax to $P_1\simeq 0.18$ and $\theta=0$ (black dot).
(b) Time evolution of $E_0(t)$; we set  $E_0(0)=0$, so the asymptotic value gives the energy dissipation. 
(c) Fourier transfrom of $E_0(t)$.
(d) Energy dissipation as a function of $\Gamma_0$: if we could set $\Gamma_0=0$ in the numerics, we could reach the limit value $-4 T_{K1}/\pi \log T_{K1}$ (horizontal line),
with $T_{K1}=z_1 \Gamma_1$, $z_1=4P_1(1-P_1)$.
}\label{fig_gutz}
\end{figure}

In Fig. \ref{fig_gutz} we show an example, where we can observe oscillations of $E_0(t)$, which were absent in the non-interacting case, Figs. \ref{fig_static}(a), \ref{fig_static_sqrt}(a).

\subsection{Time-periodic driving }
\begin{figure}[bt]
\includegraphics[width=0.45\textwidth]{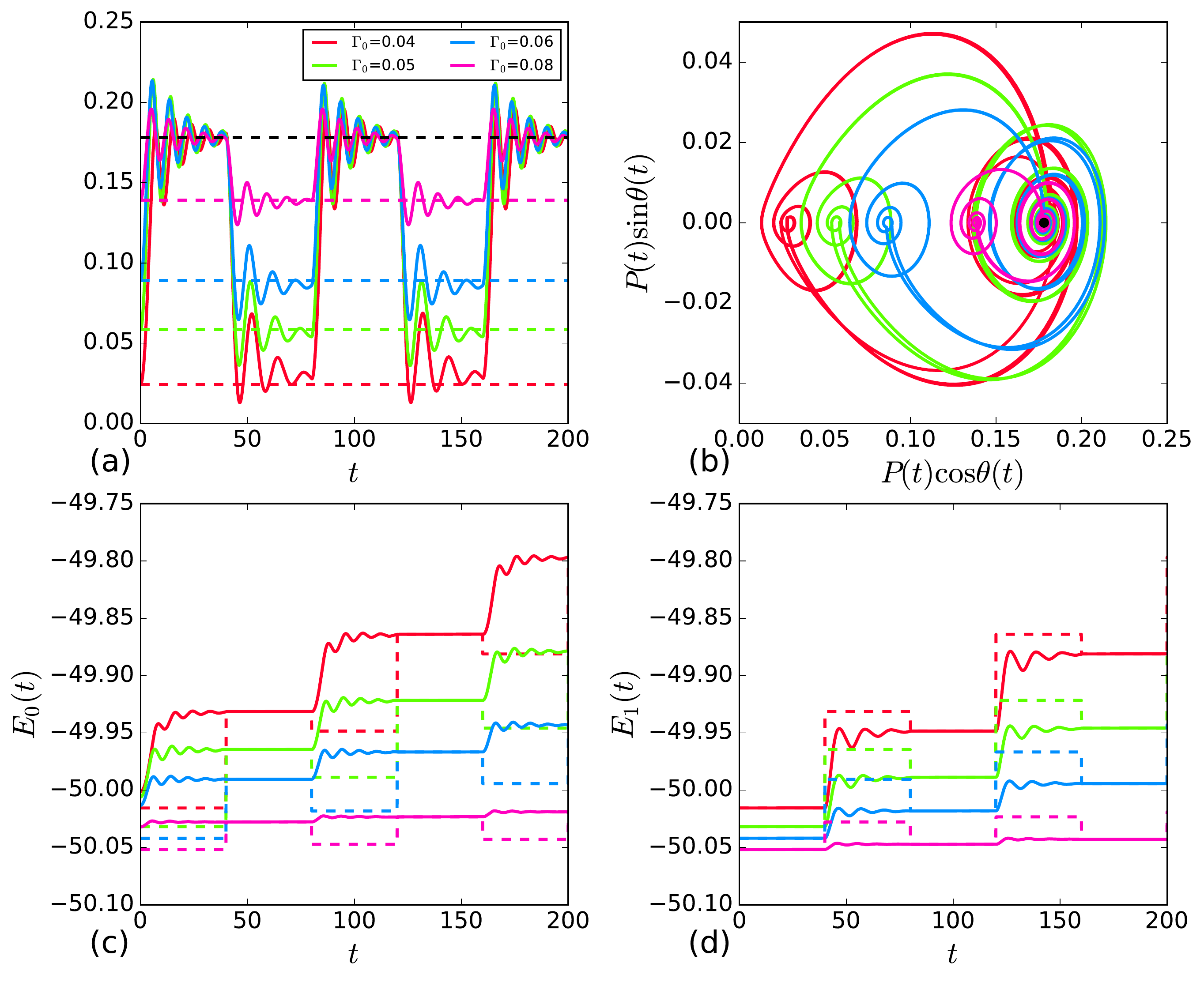}
\caption{Low frequency periodic hybridization quench in the TD-Gutzwiller model for the same parameters as in Fig. \ref{fig_gutz}, and $\tau=40$.
(a) Time evolution of $P(t)$, which, in each semiperiod, tends to relax either to $P_1$ (black line), or to $P_0$ (colored lines).
(b) Time evolution of $P(t)$ and $\theta(t)$.
(c) Time evolution of $E_0(t)$ (solid line) and $E(t)$ (dashed line).
(d) Time evolution of $E_1(t)$ (solid line) and $E(t)$ (dashed line).
}\label{fig_gutz_per1}
\end{figure}

\begin{figure}[bt]
\includegraphics[width=0.45\textwidth]{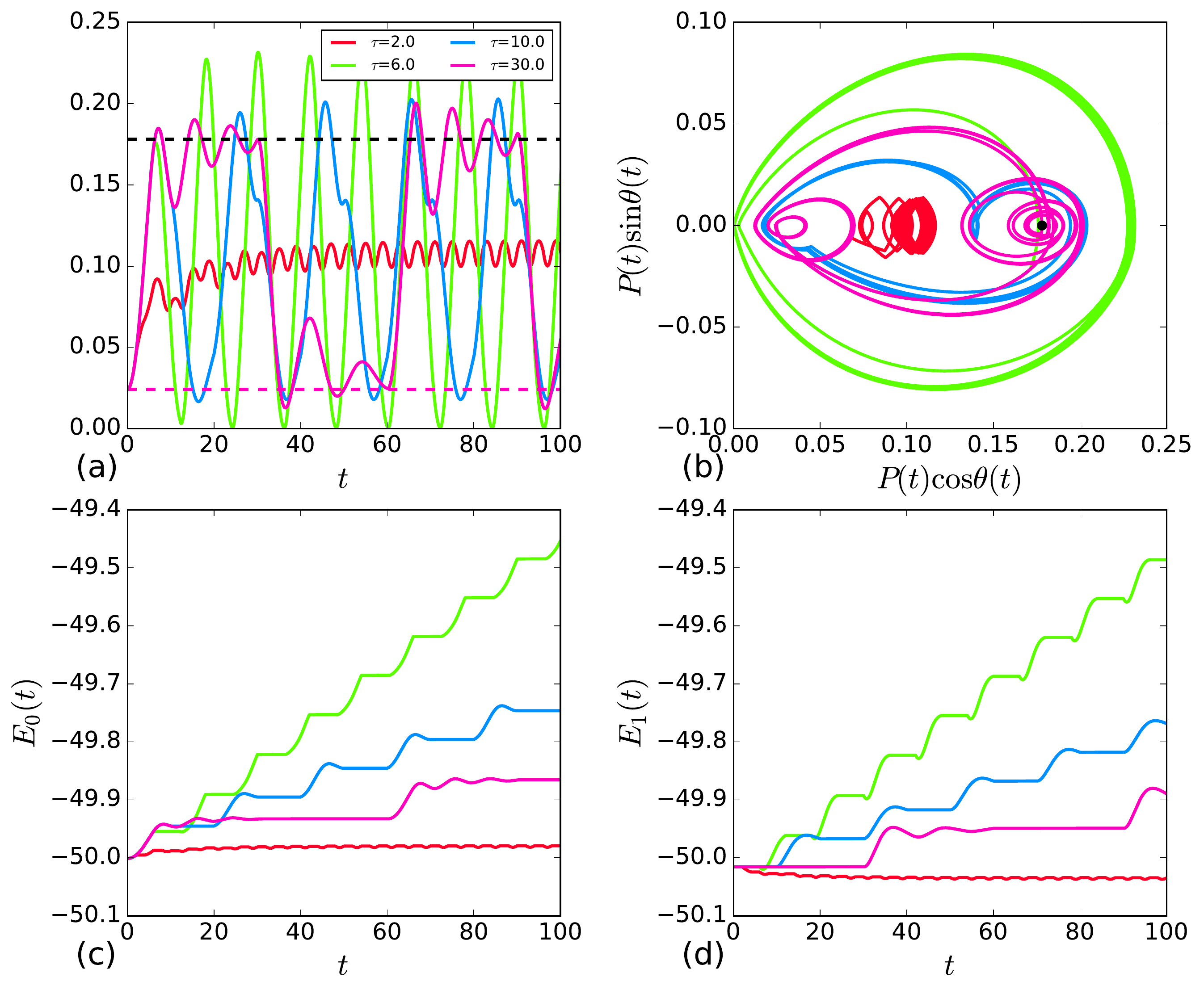}
\caption{Same as Fig. \ref{fig_gutz_per1}, but for fixed $\Gamma_0=0.04$, and different values of $\tau$.
}\label{fig_gutz_per2}
\end{figure}

Let's now consider the finite frequency case.
At low frequencies, Fig. \ref{fig_gutz_per1}, the dynamics is the one of a series of independent quenches.
As we raise the frequency, things get more complicated, Fig. \ref{fig_gutz_per2}.

\begin{figure}[bt]
\includegraphics[width=0.45\textwidth]{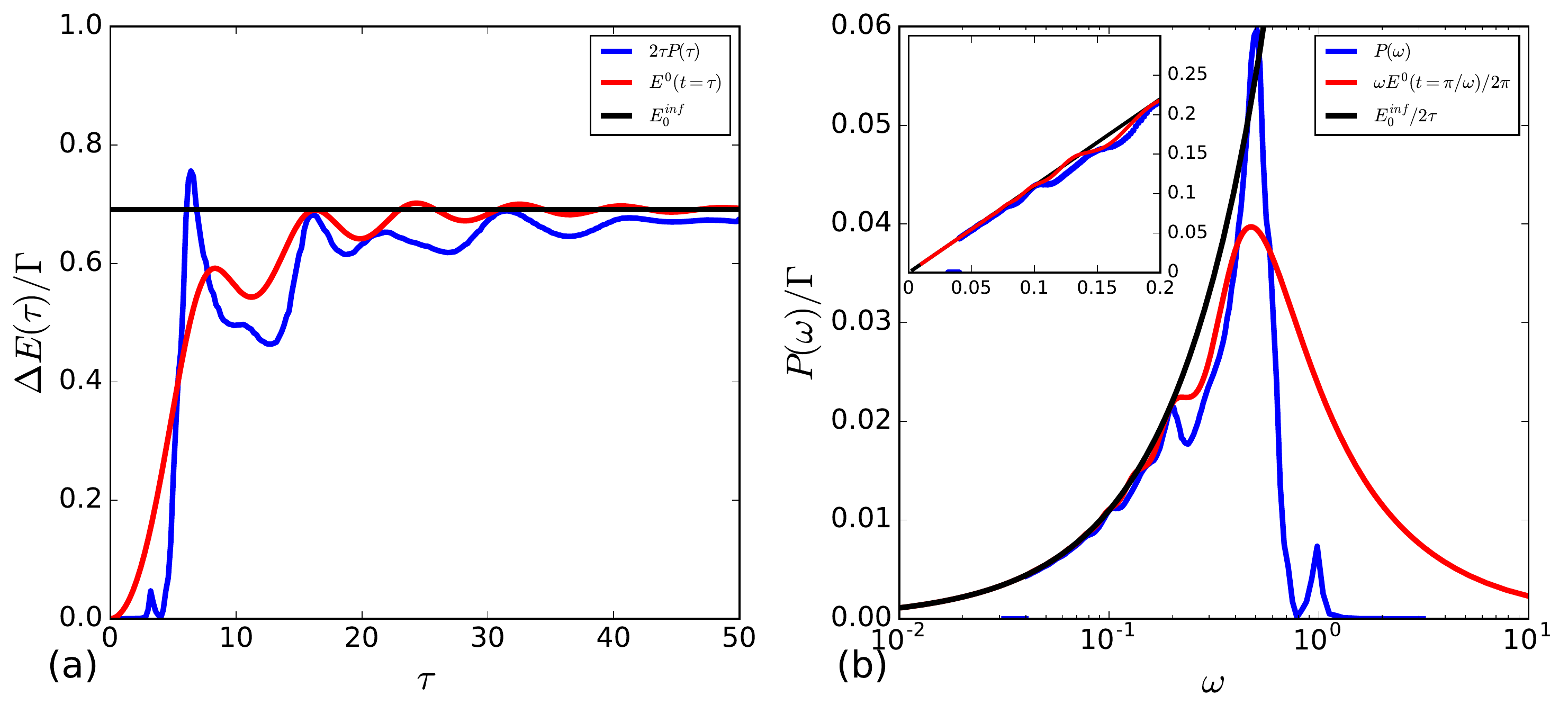}
\caption{Finite-frequency energy dissipation in the TD-Gutzwiller approximation for the same parameters as in Fig. \ref{fig_gutz}.
(a) Energy dissipation per cycle as a function of $\tau$, compared with $E_0(t=\tau)$ for a single quench.
(b) Power dissipation as a function of $\omega=\pi/\tau$.
}\label{fig_gutz_per}
\end{figure}

As in the non-interacting case, the energy dissipated per period as a function of $\tau$ follows closely $E_0(t=\tau)$ for a single quench,Fig. \ref{fig_gutz_per}.
This means that most of the results of Section \ref{sec_num_res} remain valid, except for the absence of oscillations,
that here can be observed with frequency $\sim\Gamma$.

\section{Discussion }\label{section_disc}
We have seen that, in the limit $\tau \rightarrow \infty$, or, equivalently, for a single quench, 
the RLM and the TD-Gutzwiller 
give basically the same result for the dissipation:
\be
E_{diss}=-\frac{4}{\pi}T_K\log T_K,
\ee
once we identify $T_K=\Gamma$ for the RLM (a non-interacting lorentzian level with $\epsilon_d=0$), 
and $T_K=z\Gamma$ for the Gutzwiller approximation.
In this limit $\tau \rightarrow \infty$, we also have at our disposal the results of Ref. \cite{baruselli2017},
where we computed $E_{diss}$ for a SIAM with NRG, which gives numerically exact results.
In that situation, the dissipation contained a large term proportional to $\Gamma$,
plus a smaller Kondo contribution proportional to $T_K$, albeit with logarithmic corrections not expressable  
as $\log T_K$, since the Kondo resonance is not simply a Lorentzian.

Apart from the form of the logarithmic corrections, the three metods agree in giving a dissipation proportional to $T_K$, with a somewhat large prefactor.

Things are more complicated, 
and potentially more interesting  
at finite frequency.
First of all, we do not have at our disposal results for TDNRG;
and even if we had them, TDNRG is not as good as NRG, and its results might not be completely trustworthy.
Secondly, the RLM and the the Gutzwiller approximation give different behaviors.
According to the RLM, the energy dissipated per cycle increases monotonically with $\tau$;
the power has a peak at high frequencies, on the order of the bandwidth, then decreases monotonically at lower frequencies.
According to the Gutzwiller method, the energy dissipated per cycle increases with $\tau$, but with oscillations on the order of $\Gamma$;
and the power still has a  peak at high frequencies.
The frequency-dependent behavior is qualitatively similar in both cases.

\section{Conclusions}\label{sec_concl}
In this paper we have shown two simple methods to compute the energy dissipated in switching on and off the Kondo effect, the RLM and the TD-GA. The two methods yield essentially the same result. 
The dissipation is proportional to the Kondo temperature, with logarithmic corrections which can be large.
At low frequencies the power is linear in the frequency, peaks at high frequencies, on the order of the bandwidth, then drops to zero at ultra-high frequencies.
Although at this stage we do not address any existing experiment, one may speculate that the time-dependent Kondo switching dissipation could be relevant to e.g.,  irradiated quantum dots\cite{glazman_irradiated_qd,goldstein_irradiated_qd} or other systems.


\section{Acknowledgements}
Sponsored by ERC MODPHYSFRICT Advanced Grant No. 320796
We acknowledge useful discussions with M. Fabrizio, G. Santoro and M. Goldstein.



\bibliography{biblio.bib}

\begin{thebibliography}{35}
\expandafter\ifx\csname natexlab\endcsname\relax\def\natexlab#1{#1}\fi
\expandafter\ifx\csname bibnamefont\endcsname\relax
  \def\bibnamefont#1{#1}\fi
\expandafter\ifx\csname bibfnamefont\endcsname\relax
  \def\bibfnamefont#1{#1}\fi
\expandafter\ifx\csname citenamefont\endcsname\relax
  \def\citenamefont#1{#1}\fi
\expandafter\ifx\csname url\endcsname\relax
  \def\url#1{\texttt{#1}}\fi
\expandafter\ifx\csname urlprefix\endcsname\relax\def\urlprefix{URL }\fi
\providecommand{\bibinfo}[2]{#2}
\providecommand{\eprint}[2][]{\url{#2}}

\bibitem[{\citenamefont{Baruselli et~al.}(2017)\citenamefont{Baruselli,
  Fabrizio, and Tosatti}}]{baruselli2017}
\bibinfo{author}{\bibfnamefont{P.~P.} \bibnamefont{Baruselli}},
  \bibinfo{author}{\bibfnamefont{M.}~\bibnamefont{Fabrizio}}, \bibnamefont{and}
  \bibinfo{author}{\bibfnamefont{E.}~\bibnamefont{Tosatti}},
  \bibinfo{journal}{Phys. Rev. B} \textbf{\bibinfo{volume}{96}},
  \bibinfo{pages}{075113} (\bibinfo{year}{2017}),
  \urlprefix\url{https://link.aps.org/doi/10.1103/PhysRevB.96.075113}.

\bibitem[{\citenamefont{Kaminski et~al.}(1999)\citenamefont{Kaminski, Nazarov,
  and Glazman}}]{glazman_irradiated_qd}
\bibinfo{author}{\bibfnamefont{A.}~\bibnamefont{Kaminski}},
  \bibinfo{author}{\bibfnamefont{Y.~V.} \bibnamefont{Nazarov}},
  \bibnamefont{and} \bibinfo{author}{\bibfnamefont{L.~I.}
  \bibnamefont{Glazman}}, \bibinfo{journal}{Phys. Rev. Lett.}
  \textbf{\bibinfo{volume}{83}}, \bibinfo{pages}{384} (\bibinfo{year}{1999}),
  \urlprefix\url{https://link.aps.org/doi/10.1103/PhysRevLett.83.384}.

\bibitem[{\citenamefont{Sbierski et~al.}(2013)\citenamefont{Sbierski, Hanl,
  Weichselbaum, T\"ureci, Goldstein, Glazman, von Delft, and \ifmmode
  \dot{I}\else \.{I}\fi{}mamo\ifmmode~\breve{g}\else
  \u{g}\fi{}lu}}]{goldstein_irradiated_qd}
\bibinfo{author}{\bibfnamefont{B.}~\bibnamefont{Sbierski}},
  \bibinfo{author}{\bibfnamefont{M.}~\bibnamefont{Hanl}},
  \bibinfo{author}{\bibfnamefont{A.}~\bibnamefont{Weichselbaum}},
  \bibinfo{author}{\bibfnamefont{H.~E.} \bibnamefont{T\"ureci}},
  \bibinfo{author}{\bibfnamefont{M.}~\bibnamefont{Goldstein}},
  \bibinfo{author}{\bibfnamefont{L.~I.} \bibnamefont{Glazman}},
  \bibinfo{author}{\bibfnamefont{J.}~\bibnamefont{von Delft}},
  \bibnamefont{and} \bibinfo{author}{\bibfnamefont{A.}~\bibnamefont{\ifmmode
  \dot{I}\else \.{I}\fi{}mamo\ifmmode~\breve{g}\else \u{g}\fi{}lu}},
  \bibinfo{journal}{Phys. Rev. Lett.} \textbf{\bibinfo{volume}{111}},
  \bibinfo{pages}{157402} (\bibinfo{year}{2013}),
  \urlprefix\url{https://link.aps.org/doi/10.1103/PhysRevLett.111.157402}.

\bibitem[{\citenamefont{Schir\'o and Fabrizio}(2010)}]{schiro_TDGA}
\bibinfo{author}{\bibfnamefont{M.}~\bibnamefont{Schir\'o}} \bibnamefont{and}
  \bibinfo{author}{\bibfnamefont{M.}~\bibnamefont{Fabrizio}},
  \bibinfo{journal}{Phys. Rev. Lett.} \textbf{\bibinfo{volume}{105}},
  \bibinfo{pages}{076401} (\bibinfo{year}{2010}),
  \urlprefix\url{https://link.aps.org/doi/10.1103/PhysRevLett.105.076401}.

\bibitem[{\citenamefont{Lanat\`a and Strand}(2012)}]{lanata_TDGA}
\bibinfo{author}{\bibfnamefont{N.}~\bibnamefont{Lanat\`a}} \bibnamefont{and}
  \bibinfo{author}{\bibfnamefont{H.~U.~R.} \bibnamefont{Strand}},
  \bibinfo{journal}{Phys. Rev. B} \textbf{\bibinfo{volume}{86}},
  \bibinfo{pages}{115310} (\bibinfo{year}{2012}),
  \urlprefix\url{https://link.aps.org/doi/10.1103/PhysRevB.86.115310}.

\bibitem[{\citenamefont{Schir\'o}(2012)}]{schiro2_TDGA}
\bibinfo{author}{\bibfnamefont{M.}~\bibnamefont{Schir\'o}},
  \bibinfo{journal}{Phys. Rev. B} \textbf{\bibinfo{volume}{86}},
  \bibinfo{pages}{161101} (\bibinfo{year}{2012}),
  \urlprefix\url{https://link.aps.org/doi/10.1103/PhysRevB.86.161101}.

\bibitem[{\citenamefont{Fabrizio}(2013)}]{Fabrizio_TDGA}
\bibinfo{author}{\bibfnamefont{M.}~\bibnamefont{Fabrizio}},
  \emph{\bibinfo{title}{The Out-of-Equilibrium Time-Dependent Gutzwiller
  Approximation}} (\bibinfo{publisher}{Springer Netherlands},
  \bibinfo{address}{Dordrecht}, \bibinfo{year}{2013}), pp.
  \bibinfo{pages}{247--273}, ISBN \bibinfo{isbn}{978-94-007-4984-9},
  \urlprefix\url{https://doi.org/10.1007/978-94-007-4984-9_16}.

\bibitem[{\citenamefont{Gutzwiller}(1964)}]{gutzwiller_1}
\bibinfo{author}{\bibfnamefont{M.~C.} \bibnamefont{Gutzwiller}},
  \bibinfo{journal}{Phys. Rev.} \textbf{\bibinfo{volume}{134}},
  \bibinfo{pages}{A923} (\bibinfo{year}{1964}),
  \urlprefix\url{https://link.aps.org/doi/10.1103/PhysRev.134.A923}.

\bibitem[{\citenamefont{Gutzwiller}(1965)}]{gutzwiller_2}
\bibinfo{author}{\bibfnamefont{M.~C.} \bibnamefont{Gutzwiller}},
  \bibinfo{journal}{Phys. Rev.} \textbf{\bibinfo{volume}{137}},
  \bibinfo{pages}{A1726} (\bibinfo{year}{1965}),
  \urlprefix\url{https://link.aps.org/doi/10.1103/PhysRev.137.A1726}.

\bibitem[{\citenamefont{Lanat\`a}(2010)}]{lanata_gutzwiller}
\bibinfo{author}{\bibfnamefont{N.}~\bibnamefont{Lanat\`a}},
  \bibinfo{journal}{Phys. Rev. B} \textbf{\bibinfo{volume}{82}},
  \bibinfo{pages}{195326} (\bibinfo{year}{2010}),
  \urlprefix\url{https://link.aps.org/doi/10.1103/PhysRevB.82.195326}.

\bibitem[{\citenamefont{Anders and Schiller}(2005)}]{TDNRG}
\bibinfo{author}{\bibfnamefont{F.~B.} \bibnamefont{Anders}} \bibnamefont{and}
  \bibinfo{author}{\bibfnamefont{A.}~\bibnamefont{Schiller}},
  \bibinfo{journal}{Phys. Rev. Lett.} \textbf{\bibinfo{volume}{95}},
  \bibinfo{pages}{196801} (\bibinfo{year}{2005}),
  \urlprefix\url{https://link.aps.org/doi/10.1103/PhysRevLett.95.196801}.

\bibitem[{\citenamefont{Anders and Schiller}(2006)}]{TDNRG2}
\bibinfo{author}{\bibfnamefont{F.~B.} \bibnamefont{Anders}} \bibnamefont{and}
  \bibinfo{author}{\bibfnamefont{A.}~\bibnamefont{Schiller}},
  \bibinfo{journal}{Phys. Rev. B} \textbf{\bibinfo{volume}{74}},
  \bibinfo{pages}{245113} (\bibinfo{year}{2006}),
  \urlprefix\url{https://link.aps.org/doi/10.1103/PhysRevB.74.245113}.

\bibitem[{\citenamefont{Nghiem and Costi}(2014)}]{TDNRG_costi}
\bibinfo{author}{\bibfnamefont{H.~T.~M.} \bibnamefont{Nghiem}}
  \bibnamefont{and} \bibinfo{author}{\bibfnamefont{T.~A.} \bibnamefont{Costi}},
  \bibinfo{journal}{Phys. Rev. B} \textbf{\bibinfo{volume}{89}},
  \bibinfo{pages}{075118} (\bibinfo{year}{2014}),
  \urlprefix\url{https://link.aps.org/doi/10.1103/PhysRevB.89.075118}.

\bibitem[{\citenamefont{Rosch}(2012)}]{Rosch_TDNRG}
\bibinfo{author}{\bibfnamefont{A.}~\bibnamefont{Rosch}}, \bibinfo{journal}{The
  European Physical Journal B} \textbf{\bibinfo{volume}{85}},
  \bibinfo{pages}{6} (\bibinfo{year}{2012}), ISSN \bibinfo{issn}{1434-6036},
  \urlprefix\url{https://doi.org/10.1140/epjb/e2011-20880-7}.

\bibitem[{\citenamefont{Brunner and Langreth}(1997)}]{PhysRevB.55.2578}
\bibinfo{author}{\bibfnamefont{T.}~\bibnamefont{Brunner}} \bibnamefont{and}
  \bibinfo{author}{\bibfnamefont{D.~C.} \bibnamefont{Langreth}},
  \bibinfo{journal}{Phys. Rev. B} \textbf{\bibinfo{volume}{55}},
  \bibinfo{pages}{2578} (\bibinfo{year}{1997}),
  \urlprefix\url{https://link.aps.org/doi/10.1103/PhysRevB.55.2578}.

\bibitem[{\citenamefont{Plihal and Langreth}(1998)}]{PhysRevB.58.2191}
\bibinfo{author}{\bibfnamefont{M.}~\bibnamefont{Plihal}} \bibnamefont{and}
  \bibinfo{author}{\bibfnamefont{D.~C.} \bibnamefont{Langreth}},
  \bibinfo{journal}{Phys. Rev. B} \textbf{\bibinfo{volume}{58}},
  \bibinfo{pages}{2191} (\bibinfo{year}{1998}),
  \urlprefix\url{https://link.aps.org/doi/10.1103/PhysRevB.58.2191}.

\bibitem[{\citenamefont{Plihal and Langreth}(1999)}]{PhysRevB.60.5969}
\bibinfo{author}{\bibfnamefont{M.}~\bibnamefont{Plihal}} \bibnamefont{and}
  \bibinfo{author}{\bibfnamefont{D.~C.} \bibnamefont{Langreth}},
  \bibinfo{journal}{Phys. Rev. B} \textbf{\bibinfo{volume}{60}},
  \bibinfo{pages}{5969} (\bibinfo{year}{1999}),
  \urlprefix\url{https://link.aps.org/doi/10.1103/PhysRevB.60.5969}.

\bibitem[{\citenamefont{Cohen et~al.}(2013)\citenamefont{Cohen, Gull, Reichman,
  Millis, and Rabani}}]{cohen_qmc}
\bibinfo{author}{\bibfnamefont{G.}~\bibnamefont{Cohen}},
  \bibinfo{author}{\bibfnamefont{E.}~\bibnamefont{Gull}},
  \bibinfo{author}{\bibfnamefont{D.~R.} \bibnamefont{Reichman}},
  \bibinfo{author}{\bibfnamefont{A.~J.} \bibnamefont{Millis}},
  \bibnamefont{and} \bibinfo{author}{\bibfnamefont{E.}~\bibnamefont{Rabani}},
  \bibinfo{journal}{Phys. Rev. B} \textbf{\bibinfo{volume}{87}},
  \bibinfo{pages}{195108} (\bibinfo{year}{2013}),
  \urlprefix\url{https://link.aps.org/doi/10.1103/PhysRevB.87.195108}.

\bibitem[{\citenamefont{Hackl and Kehrein}(2008)}]{kehrein_unitary_pert_theory}
\bibinfo{author}{\bibfnamefont{A.}~\bibnamefont{Hackl}} \bibnamefont{and}
  \bibinfo{author}{\bibfnamefont{S.}~\bibnamefont{Kehrein}},
  \bibinfo{journal}{Phys. Rev. B} \textbf{\bibinfo{volume}{78}},
  \bibinfo{pages}{092303} (\bibinfo{year}{2008}),
  \urlprefix\url{https://link.aps.org/doi/10.1103/PhysRevB.78.092303}.

\bibitem[{\citenamefont{Schiller and Hershfield}(1996)}]{schiller_ac}
\bibinfo{author}{\bibfnamefont{A.}~\bibnamefont{Schiller}} \bibnamefont{and}
  \bibinfo{author}{\bibfnamefont{S.}~\bibnamefont{Hershfield}},
  \bibinfo{journal}{Phys. Rev. Lett.} \textbf{\bibinfo{volume}{77}},
  \bibinfo{pages}{1821} (\bibinfo{year}{1996}),
  \urlprefix\url{https://link.aps.org/doi/10.1103/PhysRevLett.77.1821}.

\bibitem[{\citenamefont{Lobaskin and Kehrein}(2005)}]{lobaskin_prb}
\bibinfo{author}{\bibfnamefont{D.}~\bibnamefont{Lobaskin}} \bibnamefont{and}
  \bibinfo{author}{\bibfnamefont{S.}~\bibnamefont{Kehrein}},
  \bibinfo{journal}{Phys. Rev. B} \textbf{\bibinfo{volume}{71}},
  \bibinfo{pages}{193303} (\bibinfo{year}{2005}),
  \urlprefix\url{https://link.aps.org/doi/10.1103/PhysRevB.71.193303}.

\bibitem[{\citenamefont{Lobaskin and Kehrein}(2006)}]{Lobaskin2006}
\bibinfo{author}{\bibfnamefont{D.}~\bibnamefont{Lobaskin}} \bibnamefont{and}
  \bibinfo{author}{\bibfnamefont{S.}~\bibnamefont{Kehrein}},
  \bibinfo{journal}{Journal of Statistical Physics}
  \textbf{\bibinfo{volume}{123}}, \bibinfo{pages}{301} (\bibinfo{year}{2006}),
  ISSN \bibinfo{issn}{1572-9613},
  \urlprefix\url{https://doi.org/10.1007/s10955-006-9055-5}.

\bibitem[{\citenamefont{Heyl and Kehrein}(2010)}]{heyl_ackondo}
\bibinfo{author}{\bibfnamefont{M.}~\bibnamefont{Heyl}} \bibnamefont{and}
  \bibinfo{author}{\bibfnamefont{S.}~\bibnamefont{Kehrein}},
  \bibinfo{journal}{Phys. Rev. B} \textbf{\bibinfo{volume}{81}},
  \bibinfo{pages}{144301} (\bibinfo{year}{2010}),
  \urlprefix\url{https://link.aps.org/doi/10.1103/PhysRevB.81.144301}.

\bibitem[{\citenamefont{Iwahori and Kawakami}(2016)}]{iwahori_pra}
\bibinfo{author}{\bibfnamefont{K.}~\bibnamefont{Iwahori}} \bibnamefont{and}
  \bibinfo{author}{\bibfnamefont{N.}~\bibnamefont{Kawakami}},
  \bibinfo{journal}{Phys. Rev. A} \textbf{\bibinfo{volume}{94}},
  \bibinfo{pages}{063647} (\bibinfo{year}{2016}),
  \urlprefix\url{https://link.aps.org/doi/10.1103/PhysRevA.94.063647}.

\bibitem[{\citenamefont{Nordlander et~al.}(1999)\citenamefont{Nordlander,
  Pustilnik, Meir, Wingreen, and Langreth}}]{langreth_kondo}
\bibinfo{author}{\bibfnamefont{P.}~\bibnamefont{Nordlander}},
  \bibinfo{author}{\bibfnamefont{M.}~\bibnamefont{Pustilnik}},
  \bibinfo{author}{\bibfnamefont{Y.}~\bibnamefont{Meir}},
  \bibinfo{author}{\bibfnamefont{N.~S.} \bibnamefont{Wingreen}},
  \bibnamefont{and} \bibinfo{author}{\bibfnamefont{D.~C.}
  \bibnamefont{Langreth}}, \bibinfo{journal}{Phys. Rev. Lett.}
  \textbf{\bibinfo{volume}{83}}, \bibinfo{pages}{808} (\bibinfo{year}{1999}),
  \urlprefix\url{https://link.aps.org/doi/10.1103/PhysRevLett.83.808}.

\bibitem[{\citenamefont{Kaminski et~al.}(2000)\citenamefont{Kaminski, Nazarov,
  and Glazman}}]{PhysRevB.62.8154}
\bibinfo{author}{\bibfnamefont{A.}~\bibnamefont{Kaminski}},
  \bibinfo{author}{\bibfnamefont{Y.~V.} \bibnamefont{Nazarov}},
  \bibnamefont{and} \bibinfo{author}{\bibfnamefont{L.~I.}
  \bibnamefont{Glazman}}, \bibinfo{journal}{Phys. Rev. B}
  \textbf{\bibinfo{volume}{62}}, \bibinfo{pages}{8154} (\bibinfo{year}{2000}),
  \urlprefix\url{https://link.aps.org/doi/10.1103/PhysRevB.62.8154}.

\bibitem[{\citenamefont{L\'opez et~al.}(1998)\citenamefont{L\'opez, Aguado,
  Platero, and Tejedor}}]{PhysRevLett.81.4688}
\bibinfo{author}{\bibfnamefont{R.}~\bibnamefont{L\'opez}},
  \bibinfo{author}{\bibfnamefont{R.}~\bibnamefont{Aguado}},
  \bibinfo{author}{\bibfnamefont{G.}~\bibnamefont{Platero}}, \bibnamefont{and}
  \bibinfo{author}{\bibfnamefont{C.}~\bibnamefont{Tejedor}},
  \bibinfo{journal}{Phys. Rev. Lett.} \textbf{\bibinfo{volume}{81}},
  \bibinfo{pages}{4688} (\bibinfo{year}{1998}),
  \urlprefix\url{https://link.aps.org/doi/10.1103/PhysRevLett.81.4688}.

\bibitem[{\citenamefont{Goldin and Avishai}(1998)}]{PhysRevLett.81.5394}
\bibinfo{author}{\bibfnamefont{Y.}~\bibnamefont{Goldin}} \bibnamefont{and}
  \bibinfo{author}{\bibfnamefont{Y.}~\bibnamefont{Avishai}},
  \bibinfo{journal}{Phys. Rev. Lett.} \textbf{\bibinfo{volume}{81}},
  \bibinfo{pages}{5394} (\bibinfo{year}{1998}),
  \urlprefix\url{https://link.aps.org/doi/10.1103/PhysRevLett.81.5394}.

\bibitem[{\citenamefont{Goldin and Avishai}(2000)}]{PhysRevB.61.16750}
\bibinfo{author}{\bibfnamefont{Y.}~\bibnamefont{Goldin}} \bibnamefont{and}
  \bibinfo{author}{\bibfnamefont{Y.}~\bibnamefont{Avishai}},
  \bibinfo{journal}{Phys. Rev. B} \textbf{\bibinfo{volume}{61}},
  \bibinfo{pages}{16750} (\bibinfo{year}{2000}),
  \urlprefix\url{https://link.aps.org/doi/10.1103/PhysRevB.61.16750}.

\bibitem[{\citenamefont{Ng}(1996)}]{PhysRevLett.76.487}
\bibinfo{author}{\bibfnamefont{T.-K.} \bibnamefont{Ng}},
  \bibinfo{journal}{Phys. Rev. Lett.} \textbf{\bibinfo{volume}{76}},
  \bibinfo{pages}{487} (\bibinfo{year}{1996}),
  \urlprefix\url{https://link.aps.org/doi/10.1103/PhysRevLett.76.487}.

\bibitem[{\citenamefont{Hettler and Schoeller}(1995)}]{PhysRevLett.74.4907}
\bibinfo{author}{\bibfnamefont{M.~H.} \bibnamefont{Hettler}} \bibnamefont{and}
  \bibinfo{author}{\bibfnamefont{H.}~\bibnamefont{Schoeller}},
  \bibinfo{journal}{Phys. Rev. Lett.} \textbf{\bibinfo{volume}{74}},
  \bibinfo{pages}{4907} (\bibinfo{year}{1995}),
  \urlprefix\url{https://link.aps.org/doi/10.1103/PhysRevLett.74.4907}.

\bibitem[{\citenamefont{Nordlander et~al.}(2000)\citenamefont{Nordlander,
  Wingreen, Meir, and Langreth}}]{PhysRevB.61.2146}
\bibinfo{author}{\bibfnamefont{P.}~\bibnamefont{Nordlander}},
  \bibinfo{author}{\bibfnamefont{N.~S.} \bibnamefont{Wingreen}},
  \bibinfo{author}{\bibfnamefont{Y.}~\bibnamefont{Meir}}, \bibnamefont{and}
  \bibinfo{author}{\bibfnamefont{D.~C.} \bibnamefont{Langreth}},
  \bibinfo{journal}{Phys. Rev. B} \textbf{\bibinfo{volume}{61}},
  \bibinfo{pages}{2146} (\bibinfo{year}{2000}),
  \urlprefix\url{https://link.aps.org/doi/10.1103/PhysRevB.61.2146}.

\bibitem[{\citenamefont{Goker and Gedik}(2013)}]{goker_thermopower_kondo}
\bibinfo{author}{\bibfnamefont{A.}~\bibnamefont{Goker}} \bibnamefont{and}
  \bibinfo{author}{\bibfnamefont{E.}~\bibnamefont{Gedik}},
  \bibinfo{journal}{Journal of Physics: Condensed Matter}
  \textbf{\bibinfo{volume}{25}}, \bibinfo{pages}{365301}
  (\bibinfo{year}{2013}),
  \urlprefix\url{http://stacks.iop.org/0953-8984/25/i=36/a=365301}.

\bibitem[{\citenamefont{{Zhou, Yi} and {Ng, Tai-Kai}}(2009)}]{ng_edge}
\bibinfo{author}{\bibnamefont{{Zhou, Yi}}} \bibnamefont{and}
  \bibinfo{author}{\bibnamefont{{Ng, Tai-Kai}}}, \bibinfo{journal}{EPL}
  \textbf{\bibinfo{volume}{86}}, \bibinfo{pages}{17004} (\bibinfo{year}{2009}),
  \urlprefix\url{https://doi.org/10.1209/0295-5075/86/17004}.

\bibitem[{\citenamefont{Misra and Sudarshan}(1977)}]{quantum_zeno}
\bibinfo{author}{\bibfnamefont{B.}~\bibnamefont{Misra}} \bibnamefont{and}
  \bibinfo{author}{\bibfnamefont{E.~C.~G.} \bibnamefont{Sudarshan}},
  \bibinfo{journal}{Journal of Mathematical Physics}
  \textbf{\bibinfo{volume}{18}}, \bibinfo{pages}{756} (\bibinfo{year}{1977}),
  \eprint{https://doi.org/10.1063/1.523304},
  \urlprefix\url{https://doi.org/10.1063/1.523304}.

\end{thebibliography}

\end{document}